\documentclass[floatfix,twocolumn,showpacs,preprintnumbers,amsmath,amssymb,pra,superscriptaddress,longbibliography]{revtex4-1}
\usepackage{color}
\usepackage[usenames,dvipsnames,svgnames,table]{xcolor}
\usepackage[colorlinks=true,linkcolor=blue,urlcolor=blue,citecolor=blue]{hyperref}
\usepackage{mathtools}
\usepackage{graphicx}
\usepackage{dcolumn}
\usepackage{array}
\usepackage{lipsum}
\usepackage{bm}
\usepackage{subfigure}
\usepackage{amssymb}
\usepackage{multirow}
\usepackage{tabularx}
\usepackage{amsmath}
\usepackage{braket}
\usepackage{csquotes}
\graphicspath{{plots/}}
 \usepackage{lipsum}
\usepackage{mathrsfs}

\newcommand{\beq}{\begin{equation}}
\newcommand{\eeq}{\end{equation}}
\newcommand{\bea}{\begin{eqnarray}}
\newcommand{\eea}{\end{eqnarray}}

\begin{document}
\title{
Spin-resolved density response of the warm dense electron gas
}

\author{Tobias Dornheim}
\email{t.dornheim@hzdr.de}

\affiliation{Center for Advanced Systems Understanding (CASUS), D-02826 G\"orlitz, Germany}
\affiliation{Helmholtz-Zentrum Dresden-Rossendorf (HZDR), D-01328 Dresden, Germany}

\author{Jan Vorberger}
\affiliation{Helmholtz-Zentrum Dresden-Rossendorf (HZDR), D-01328 Dresden, Germany}

\author{Zhandos A.~Moldabekov}

\affiliation{Center for Advanced Systems Understanding (CASUS), D-02826 G\"orlitz, Germany}
\affiliation{Helmholtz-Zentrum Dresden-Rossendorf (HZDR), D-01328 Dresden, Germany}

\author{Panagiotis Tolias}
\affiliation{Space and Plasma Physics, Royal Institute of Technology (KTH), Stockholm, SE-100 44, Sweden}

\begin{abstract}
We present extensive new \emph{ab initio} path integral Monte Carlo (PIMC) results for the spin-resolved density response of the uniform electron gas (UEG) at warm dense matter conditions. This allows us to unambiguously assess the accuracy of previous theoretical approximations, thereby providing valuable new insights for the future development of dielectric schemes. From a physical perspective, we observe a nontrivial manifestation of an effective electron--electron attraction that emerges in the spin-offdiagonal static density response function at strong coupling, $r_s\gtrsim5$. All PIMC results are freely available online and can be used to benchmark new approximations and simulation schemes.
\end{abstract}
\maketitle

\section{Introduction\label{sec:introduction}}

The uniform electron gas (UEG)~\cite{loos,review,quantum_theory}, also known as \emph{jellium} or \emph{quantum one-component plasma (OCP)} in the literature, constitutes one of the most fundamental model systems in physics, quantum chemistry, and related disciplines. Having originally been introduced as a simple model description of valence electrons in metals~\cite{mahan1990many}, the UEG offers a wealth of exciting physical effects such as Wigner crystallization~\cite{Wigner_PhysRev_1934,PhysRevB.69.085116,Azadi_Wigner_2022}, a \emph{roton feature} in the dynamic structure factor $S(\mathbf{q},\omega)$~\cite{dornheim_dynamic,Dornheim_Nature_2022,Takada_PRB_2016}, and an effective electronic attraction without the mediation of the ionic component~\cite{Takada_PRB_1993,Dornheim_Force_2022}. In fact, the UEG has been of pivotal importance for a number of groundbreaking developments such as the Fermi liquid theory~\cite{pines,Eich_PRB_2017} and the BCS theory of superconductivity~\cite{Bardeen_PhysRev_1957}. Furthermore, the possibly unrivaled success of density functional theory (DFT) regarding the description of real materials~\cite{Jones_RMP_2015} has been facilitated by highly accurate numerical results for the exchange--correlation properties of the UEG~\cite{Ceperley_Alder_PRL_1980}, which have been employed as input for the construction of widely used parametrizations~\cite{Perdew_Zunger_PRB_1981,vwn,Perdew_Wang_PRB_1992,farid}.

Over the recent years, an additional interest has emerged in the properties of matter at extreme densities and temperatures. Such \emph{warm dense matter} (WDM) conditions are rather ubiquitous throughout our universe~\cite{fortov_review}, and naturally occur in a number of dense astrophysical objects, like the interior of giant planets~\cite{Benuzzi_Mounaix_2014,Militzer_2008}, brown dwarfs~\cite{becker,saumon1}, and the outer layer of neutron stars~\cite{Haensel,neutron_star_envelopes}. Moreover, WDM plays an important role in cutting-edge technological applications such as inertial confinement fusion~\cite{hu_ICF,Betti2016}, novel material discovery and fabrication~\cite{Kraus2016,Kraus2017,Lazicki2021}, and hot-electron chemistry~\cite{Brongersma2015}. A topical overview of different experimental techniques for the study of WDM has been given by Falk~\cite{falk_wdm}.

Naturally, the accurate theoretical WDM description requires the extension of previous UEG studies to the relevant density-temperature regime. With the exception of the symmetric spin-unpolarized case, the finite temperature UEG should be considered as a two-component system of spin-up (designation \enquote{u}) and spin-down (designation \enquote{d}) electrons. Thus, its thermodynamic state requires three state variables to be fully specified. The following three dimensionless parameters are typically employed: (I) The Wigner-Seitz radius or quantum coupling parameter $r_{\mathrm{s}}=\overline{a}/a_\textnormal{B}$ with $\overline{a}$ and $a_\textnormal{B}$ being the mean distance to the nearest neighbour and the first Bohr radius. (II) The degeneracy temperature $\theta=k_\textnormal{B}T/E_\textnormal{F}$ with $T$ the temperature and $E_\textnormal{F}=\hbar^2(k_{\mathrm{F}}^{\mathrm{u}})^2/(2m)$ being the  Fermi energy defined with respect to the Fermi wave vector of the spin-up electrons $k_{\mathrm{F}}^{\mathrm{u}}=(6\pi^2n_{\mathrm{u}})^{1/3}$ (under the convention that the spin-up species is the highest density species~\cite{quantum_theory}). (III) The spin polarization $\xi=(n_{\mathrm{u}}-n_{\mathrm{d}})/n$, $0\leq\xi\leq1$, with $n_{\mathrm{u}}$ ($n_{\mathrm{d}}$) the density of spin-up (spin-down) electrons, and with $n=n_{\mathrm{u}}+n_{\mathrm{d}}$ the total electron density. Specifically, in WDM conditions, $r_{\mathrm{s}}\sim\theta\sim1$~\cite{wdm_book,new_POP,Ott2018}, which implies that there exists no small parameter to perform an expansion around~\cite{wdm_book}. In other words, the rigorous theoretical WDM description must cover the highly nontrivial interplay of moderate Coulomb correlations with strong thermal excitations and quantum effects such as diffraction and Pauli blocking. 

These challenges have sparked a recent surge of activity in the description of the UEG at WDM conditions~\cite{Dutta_2013,stls2,Brown_PRL_2013,Brown_PRB_2013,ksdt,status,Dornheim_JCP_2015,Malone_JCP_2015,Filinov_PRE_2015,Schoof_PRL_2015,Groth_PRB_2016,Malone_PRL_2016,Dornheim_PRB_2016,dornheim_prl,Dornheim_POP_2017,Kas_PRL_2017,dornheim_pre,groth_jcp,Tanaka_CPP_2017,groth_prl,tanaka_hnc,review,arora,Kas_PRB_2019,dornheim_HEDP,dornheim_ML,Dornheim_PRB_2021,Dornheim_PRL_2020_ESA,Yilmaz_JCP_2020,Ara_POP_2021,Joonho_JCP_2021,Filinov_CPP_2021,Hunger_PRE_2021,Dornheim_PRB_nk_2021,https://doi.org/10.48550/arxiv.2202.02736}, which has led to the first accurate parametrizations of the exchange--correlation free energy valid in the entire WDM range~\cite{ksdt,groth_prl}; see Refs.~\cite{Dornheim_POP_2017,review,status} for overviews of different aspects of these developments. More specific, such parametrizations constitute the basis for thermal DFT \cite{Mermin_DFT_1965} WDM simulations that consistently treat the interplay of thermal and exchange--correlation effects~\cite{karasiev_importance,kushal} on the level of the local density approximation. 

While being an important milestone in the right direction, a more thorough WDM theory requires additional information. In this context, a key concept is the UEG response to external perturbations; information that is indispensable for the interpretation of state-of-the-art X-ray Thomson scattering experiments~\cite{siegfried_review,kraus_xrts,Dornheim_PRL_2020_ESA}, which constitutes the de-facto diagnostic at these extreme conditions. Other applications include the estimation of the electronic stopping power~\cite{Moldabekov_PRE_2020}, construction of effective ion--ion potentials~\cite{ceperley_potential,zhandos1,zhandos2} and development of advanced exchange--correlation functionals for DFT~\cite{Patrick_JCP_2015,pribram}. 

Consequently, the UEG density response has been extensively studied using semi-analytical theories~\cite{IIT,stls,stls2,stls_original,vs_original,farid,Tolias_JCP_2021,castello2021classical} and numerical methods~\cite{dornheim_pre,groth_jcp,moroni,moroni2,bowen2,dornheim_ML,dornheim_HEDP}. In the ground state, the first accurate results have been obtained by Moroni \emph{et al.}~\cite{moroni,moroni2}, who carried out quantum Monte Carlo (QMC) calculations of a harmonically perturbed UEG (see also Ref.~\cite{bowen2}). Subsequently, these data were used as input for the widely used parametrization by Corradini \emph{et al.}~\cite{cdop}. Recently, Chen and Haule~\cite{Chen2019} presented new diagrammatic QMC results at metallic densities, which substantiated previous results at $T=0$.

The first finite temperature QMC results have been presented by Dornheim, Groth and co-workers~\cite{dornheim_pre,groth_jcp}, although the studies were limited to a few temperature--density combinations. A more extensive data set was presented in Ref.~\cite{dornheim_ML} based on imaginary-time correlation functions~\cite{Dornheim_JCP_ITCF_2021}, which allow estimates of the full wavenumber static density response dependence from a single simulation of the unperturbed system. In combination with the ground state Corradini \emph{et al.} parametrization~\cite{cdop}, the data were used to train a neural-network representation of the static local field correction [cf.~Eq.~(\ref{eq:resolved_LFC}) below], which covers the entire relevant range of WDM parameters. For completeness, we note that highly accurate UEG results at WDM conditions have even become available for dynamic properties such as $S(\mathbf{q},\omega)$~\cite{dornheim_dynamic,dynamic_folgepaper,Hamann_PRB_2020} and for the nonlinear density response~\cite{Dornheim_PRL_2020,Dornheim_PRR_2021,Dornheim_JCP_ITCF_2021,Dornheim_JPSJ_2021,Dornheim_CPP_2022}.

In the present work, we extend previous studies of the linear response of the warm dense UEG by presenting the first, highly accurate \emph{ab initio} path integral Monte Carlo (PIMC)~\cite{cep,dornheim_sign_problem} results for the spin-resolved components of density response function, and the static local field correction. First and foremost, we note that this constitutes an indispensable basis for the study of WDM in the presence of an external magnetic field~\cite{Dornheim_PRE_2021, PhysRevB.105.035134}, for example in the vicinity of a neutron star~\cite{Haensel}. Moreover, we gain new insights into dielectric theories such as the celebrated scheme by Singwi \emph{et al.}~\cite{stls_original,stls} (STLS), which is often remarkably accurate in the description of the full electronic density response, but considerably less accurate regarding the individual spin-resolved components. From a physical perspective, our analysis reveals the interesting and nontrivial manifestation of the recently reported effective electronic attraction within the UEG~\cite{Dornheim_Force_2022} onto the spin-offdiagonal components of the static density response function. All the PIMC results are freely available online~\cite{repo} and can be used to benchmark approximations and new simulation schemes.

The article is organized as follows: In Sec.~\ref{sec:theory}, we introduce the required theoretical background, including the generalized Hamiltonian and its relation to spin-resolved density responses (\ref{sec:Hamiltonian}), the concept of imaginary-time intermediate scattering functions~(\ref{sec:ITCF}), and the definition of spin-resolved local field corrections (\ref{sec:LFC}).
Sec.~\ref{sec:results} is devoted to our extensive new simulation results, starting with a brief discussion of the intermediate scattering function evaluated at imaginary-time arguments~(\ref{sec:ITCF_results}). In Sec.~\ref{sec:spin}, we discuss in detail the spin-resolved density response of the UEG both at a metallic density (\ref{sec:spin}) and in the strongly coupled electron liquid regime (\ref{sec:liquid}). Finally, we consider the density response at the intermediate spin-polarization of $\xi=1/3$ (\ref{sec:intermediate}). The paper is concluded by a brief summary and outlook in Sec.~\ref{sec:summary}.

\section{Theory\label{sec:theory}}

\subsection{Hamiltonian perturbation and spin-resolved static density responses\label{sec:Hamiltonian}}

Throughout this work, we consider a Hamiltonian operator of the form, 
\begin{widetext}
\begin{eqnarray}\label{eq:Hamiltonian}
\hat{H} = \sum_{s\in\{\mathrm{u,d}\}}\hat{K}_s + \frac{1}{2}\sum_{s,t\in\{\mathrm{u,d}\}} \hat{W}_{s,t} + \sum_{s\in\{\mathrm{u,d}\}} 2A_s \sum_{k=1}^{N_s}\textnormal{cos}\left(\mathbf{q}\cdot\hat{\mathbf{r}}_k\right)\ ,
\end{eqnarray}
\end{widetext}
with $\hat{K}_s$ being the kinetic energy operator of the species $s$, and $\hat{W}_{s,t}$ taking into account the full two-body interaction between all particles from species $s$ and $t$. In the UEG case, the interaction is given by the usual Ewald sum as it has been introduced in detail e.g.~in Ref.~\cite{Fraser_PRB_1996}. The final term in Eq.~(\ref{eq:Hamiltonian}) corresponds to an external static harmonic perturbation~\cite{moroni,moroni2,bowen2,Dornheim_PRL_2020,dornheim_pre,groth_jcp,Bohme_PRL_2022} with a species-dependent amplitude $A_s$. 

Within linear response theory, \emph{i.e.}, in the limit of infinitesimal perturbations, the static density response of species $s$ is given by~\cite{IIT}
\begin{eqnarray}\label{eq:LRT}
\delta\braket{\hat{\rho}_s(\mathbf{k})} = \sum_{t\in\{\mathrm{u,d}\}} \chi_{st}(\mathbf{k}) A_t \delta_{\mathbf{k},\mathbf{q}}\ ,
\end{eqnarray}
with $\chi_{st}(\mathbf{k})$ the spin-resolved density response function that describes the impact of a perturbation of species $t$ on the (unperturbed) species $s$ [i.e., $A_t>0$ and $A_s=0$ in Eq.~(\ref{eq:Hamiltonian})]. In that case, Eq.~(\ref{eq:LRT}) can be simplified to
\begin{eqnarray}\label{eq:delta_rho}
\delta\braket{\hat{\rho}_s(\mathbf{k})} = \chi_{st}(\mathbf{k})A_t\delta_{\mathbf{k},\mathbf{q}}\ .
\end{eqnarray}
Eq.(\ref{eq:LRT}) directly implies that the density response attains only nonzero values at the wave vector of the original perturbation, $\mathbf{k}=\mathbf{q}$. As a consequence, the excitation of higher-order harmonics constitutes a purely nonlinear phenomenon, see Refs.~\cite{Dornheim_PRR_2021,Dornheim_JCP_ITCF_2021, Dornheim_JPSJ_2021,Dornheim_CPP_2021,Dornheim_CPP_2022} for a recent investigation of such effects in the UEG.
The estimation of the LHS.~of Eq.~(\ref{eq:delta_rho}) is straightforward in our PIMC simulations,
\begin{widetext}
\begin{eqnarray}\label{eq:delta_rho2}
 \delta\braket{\hat{\rho}_s(\mathbf{k})} &=& \underbrace{\frac{1}{V} \left< \sum_{l=1}^{N_s} e^{-i\mathbf{k}\cdot\hat{\mathbf{r}}_l} \right>_{\{A_t\},\mathbf{q}}}_{\braket{\hat{\rho}_s(\mathbf{k})}}\  -\quad \underbrace{\frac{1}{V} \left< \sum_{l=1}^{N_s} e^{-i\mathbf{k}\cdot\hat{\mathbf{r}}_l} \right>_{\{A_t=0\},\mathbf{q}}}_{=0}\ .
\end{eqnarray}
\end{widetext}
We point out that the second term on the RHS.~of Eq.~(\ref{eq:delta_rho2}) corresponds to the expectation value of the density in reciprocal space in the unperturbed limit, which vanishes in the case of a uniform system~\cite{quantum_theory}. 

From a theoretical perspective, within the polarization potential approach, it is very convenient to express the density response of species $s$ in terms of the density response of an ideal (non-interacting) Fermi gas $\chi_{ss}^{(0)}(\mathbf{k})$, which leads to~\cite{IIT}
\begin{widetext}
\begin{eqnarray}\label{eq:resolved_LFC}
\delta\braket{\hat{\rho}_s(\mathbf{k})} = \chi_{ss}^{(0)}(\mathbf{k})\left(
A_s\delta_{\mathbf{k},\mathbf{q}} + \frac{4\pi}{k^2}\sum_{t\in\{\mathrm{u,d}\}} [ 1-G_{st}(\mathbf{k})] \delta\braket{\hat{\rho}_t(\mathbf{k})} 
\right)\ .
\end{eqnarray}
\end{widetext}
For completeness, we note that there exist no cross terms in the ideal Fermi case, \emph{i.e.,} $\chi_{st}^{(0)}(\mathbf{k})=\delta_{st}\chi_{st}^{(0)}(\mathbf{k})$, as different non-interacting species cannot be correlated with each other. The complete wave-number resolved information about electronic exchange--correlation effects is now fully encoded into the static local field corrections (LFCs) $G_{st}(\mathbf{k})$, which are a-priori unknown. In the present investigation, we present accurate results for all the spin-resolved LFC components that we extract from our new \emph{ab initio} PIMC data for the static density response, see Eqs.~(\ref{eq:G_components},\ref{eq:LFC}) below.

\subsection{Spin-resolved imaginary-time intermediate scattering functions\label{sec:ITCF}}

In the previous section, we have introduced a straightforward way to obtain the spin-resolved density response of the UEG from PIMC simulations of the harmonically perturbed system. While being formally exact, this procedure requires individual PIMC simulations for multiple values of the respective perturbation amplitude $A_t$ to obtain the density response $\chi_{st}(\mathbf{k})$ \emph{for a singe wave vector at one specific density--temperature combination}~\cite{dornheim_pre,groth_jcp}. In practice, one can never really know when the perturbation amplitude is sufficiently small for linear response theory to be truly valid and one cannot employ too small perturbation amplitudes due to noise issues.
A more convenient way is given by the estimation of imaginary-time correlation functions, which give us access to the full $\mathbf{k}$-dependence of the density response from a single simulation of the unperturbed system~\cite{Dornheim_JCP_ITCF_2021}. For example, the total, spin-unresolved imaginary-time version of the intermediate scattering function (IT-ISF) is given by
\begin{eqnarray}\label{eq:define_F}
F_\textnormal{tot}(\mathbf{q},\tau) = \frac{1}{N} \braket{\hat{n}_\textnormal{tot}(\mathbf{q},\tau)\hat{n}_\textnormal{tot}(-\mathbf{q},0)}_0\ ,
\end{eqnarray}
with the definition of the (unnormalized) single-particle density operator
\begin{eqnarray}
\hat{n}_\textnormal{tot}(\mathbf{q},\tau) = \sum_{k=1}^{N}e^{-i\mathbf{q}\cdot\hat{\mathbf{r}}_{k,\tau}}\ ;
\end{eqnarray}
here $\hat{\mathbf{r}}_{k,\tau}$ denotes the coordinate of particle $k$ at the imaginary time $\tau$. The utility of Eq.~(\ref{eq:define_F}) is demonstrated by its relation to the dynamic structure factor $S(\mathbf{q},\omega)$, which reads as
\begin{eqnarray}\label{eq:Laplace}
F_\textnormal{tot}(\mathbf{q},\tau) = \int_{-\infty}^\infty \textnormal{d}\omega\ S(\mathbf{q},\omega)\ e^{-\tau\omega}\ .
\end{eqnarray}
This constitutes the basis for an \emph{analytic continuation}, \emph{i.e.}, the numerical solution of Eq.~(\ref{eq:Laplace}) for $S(\mathbf{q},\omega)$. Such an \emph{inverse Laplace transform} is a notoriously hard problem, that is, unfortunately, ill-conditioned with respect to the Monte Carlo error bars~\cite{JARRELL1996133,Goulko_PRB_2017}. For the UEG, this problem has recently been overcome by Dornheim and co-workers~\cite{dornheim_dynamic,dynamic_folgepaper,Hamann_PRB_2020,Dornheim_PRE_2020} who have presented the first highly accurate results for $S(\mathbf{q},\omega)$ (and related properties) based on the stochastic sampling of the frequency-dependent LFC $G(\mathbf{q},\omega)$.

In the context of the present work, the central utility of $F(\mathbf{q},\tau)$ is captured in the imaginary-time version of the fluctuation--dissipation theorem~\cite{bowen2},
\begin{eqnarray}\label{eq:integrate}
\chi_\textnormal{tot}(\mathbf{q}) &=& - \frac{N}{V} \int_0^\beta \textnormal{d}\tau\ F(\mathbf{q},\tau)\ ,
\end{eqnarray}
which gives us direct access to the full static density response of the UEG. In addition, it is straightforward to decompose Eq.~(\ref{eq:integrate}) into separate, spin-resolved contributions
\begin{widetext}
\begin{eqnarray}\label{eq:integrate_spin}
\chi_\textnormal{tot}(\mathbf{q})
&=& - \frac{N}{V}  \int_0^\beta \textnormal{d}\tau\ \left( \underbrace{\frac{1}{N}\braket{\hat{n}_{\mathrm{u}}(\mathbf{q},\tau)\hat{n}_{\mathrm{u}}(-\mathbf{q},0)}_0}_{F_{\mathrm{uu}}(\mathbf{q},\tau)}
+\underbrace{\frac{1}{N}\braket{\hat{n}_{\mathrm{d}}(\mathbf{q},\tau)\hat{n}_{\mathrm{d}}(-\mathbf{q},0)}_0}_{F_{\mathrm{dd}}(\mathbf{q},\tau)}
+ 2 \underbrace{\frac{1}{N}\braket{\hat{n}_{\mathrm{u}}(\mathbf{q},\tau)\hat{n}_{\mathrm{d}}(-\mathbf{q},0)}_0}_{F_{\mathrm{ud}}(\mathbf{q},\tau)=F_{\mathrm{du}}(\mathbf{q},\tau)}
\right)\ ,
\end{eqnarray}
\end{widetext}
where the spin-resolved single-particle density operator in reciprocal imaginary time space is defined as
\begin{eqnarray}\label{eq:spin_resolved_density}
\hat{n}_s(\mathbf{q},\tau) = \sum_{k=1}^{N_s}e^{-i\mathbf{q}\cdot\hat{\mathbf{r}}_{k,\tau}}\ .
\end{eqnarray}
Splitting the integral in Eq.~(\ref{eq:integrate_spin}) into its individual constituents leads to the relation
\begin{eqnarray}\label{eq:Chi_spin}
\chi_{st}(\mathbf{q}) = - \frac{N}{V} \int_0^\beta \textnormal{d}\mathbf{\tau}\ F_{st}(\mathbf{q},\tau)\ ,
\end{eqnarray}
which is employed throughout this work. Furthermore, the spin-resolved static structure factor is defined as the $\tau\to0$ limit of the corresponding IT-ISF,
\begin{eqnarray}\label{eq:S_st}
\lim_{\tau\to0}F_{st}(\mathbf{q},\tau) = S_{st}(\mathbf{q})\ .
\end{eqnarray}

\subsection{Spin-resolved static local field corrections\label{sec:LFC}}

In previous works~\cite{dornheim_ML,dornheim_HEDP,dornheim_electron_liquid,review,dynamic_folgepaper}, we considered the total LFC of the spin-unpolarized UEG, which describes the impact of all electronic exchange--correlation effects on the density response of the full electronic density,
\begin{eqnarray}\label{eq:define_LFC}
\chi_\textnormal{tot}(\mathbf{q}) = \frac{\chi^{(0)}_\textnormal{tot}(\mathbf{q})}{1-\frac{4\pi}{q^2}\left[1-G_\textnormal{tot}(\mathbf{q})\right]\chi^{(0)}_\textnormal{tot}(\mathbf{q})}
\end{eqnarray}
It is apparent from Eq.~(\ref{eq:integrate_spin}) that the connection between the total density response functions and the spin-resolved density response functions is simply given by~\cite{quantum_theory,SingwiTosi_Review}
\begin{eqnarray}
\chi_\textnormal{tot}(\mathbf{q}) &=& \sum_{s,t\in\{\mathrm{u,d}\}}\chi_{st}(\mathbf{q})\,, \\ \nonumber
\chi_\textnormal{tot}^{(0)}(\mathbf{q}) &=& \sum_{s\in\{\mathrm{u,d}\}}\chi_{ss}^{(0)}(\mathbf{q})\,,
\end{eqnarray}
where we note the reciprocity relation $\chi_{\mathrm{ud}}(\mathbf{q})=\chi_{\mathrm{du}}(\mathbf{q})$ that is valid regardless of the UEG spin polarization $\xi$. Within linear response theory and the polarization potential approach for a two-component system, employing a tilde operator that inverts the spin state ($\widetilde{\mathrm{d}}=\mathrm{u}$, $\widetilde{\mathrm{u}}=\mathrm{d}$), the spin-resolved generalization of Eq.~(\ref{eq:define_LFC}) can be compactly written as~\cite{SingwiTosi_Review,IIT}
\begin{eqnarray}\label{eq:chi_components}
&\chi_{st}(\boldsymbol{q})=\chi_{tt}^{(0)}(\boldsymbol{q})\displaystyle\frac{\delta_{st}+\displaystyle(-1)^{\delta_{st}}\Gamma_{\widetilde{t}\widetilde{s}}(\boldsymbol{q})}{D(\boldsymbol{q})}\,,
\end{eqnarray}
with the definitions
\begin{align}
&D(\mathbf{q})= \left[1-\Gamma_{\mathrm{uu}}(\mathbf{q})\right]\left[1-\Gamma_{\mathrm{dd}}(\mathbf{q})\right]-\Gamma_{\mathrm{ud}}(\mathbf{q})\Gamma_{\mathrm{du}}(\mathbf{q})\,,\\
&\Gamma_{st}(\mathbf{q})=\frac{4\pi}{q^2}\left[1-G_{st}(\mathbf{q})\right]\chi_{ss}^{(0)}(\mathbf{q})\,.
\end{align}
Naturally, Eq.~(\ref{eq:chi_components}) can be straightforwardly inverted for the individual spin-resolved LFCs leading to
\begin{eqnarray}\label{eq:G_components}
\Gamma_{st}(\boldsymbol{q})=\delta_{st}+\displaystyle(-1)^{\delta_{st}}\displaystyle\frac{\chi_{ss}^{(0)}(\boldsymbol{q})\chi_{\widetilde{t}\widetilde{s}}(\boldsymbol{q})}{\mathrm{det}[\chi_{st}(\boldsymbol{q})]}\,,
\end{eqnarray}
where $\mathrm{det}[\chi_{st}(\boldsymbol{q})]$ denotes the determinant of the spin resolved density response matrix. The final result for the LFC of an arbitrary spin-component is then given by
\begin{eqnarray}\label{eq:LFC}
G_{st}(\mathbf{q}) &=& 1 - \frac{q^2}{4\pi}\frac{\Gamma_{st}(\mathbf{q})}{\chi^{(0)}_{ss}(\mathbf{q})}\,,
\end{eqnarray}
where we note the reciprocity relation $G_{\mathrm{ud}}(\mathbf{q})=G_{\mathrm{du}}(\mathbf{q})$ that is valid regardless of the UEG spin polarization $\xi$. In practice, we perform extensive new PIMC simulations of the unperturbed UEG in order to estimate $F_{st}(\mathbf{q},\tau)$ over the entire relevant range of wavenumbers. These results are then inserted into Eq.~(\ref{eq:Chi_spin}) to compute $\chi_{st}(\mathbf{q})$, which is used to calculate $\Gamma_{st}(\mathbf{q})$ via Eq.~(\ref{eq:G_components}), which, in turn, is used to estimate $G_{st}(\mathbf{q})$ via Eq.~(\ref{eq:LFC}).

It is worth pointing out that the longitudinal spin-spin response function $\chi_{\mathrm{S}}$, which gives the response of the $z$-component of the spin density to a static magnetic field perturbation parallel to the $z$-axis, is given by~\cite{quantum_theory}
\begin{eqnarray*}
\chi_\textnormal{S}(\mathbf{q})= \sum_{s,t}\mathrm{sgn}(s)\mathrm{sgn}(t)\chi_{st}(\mathbf{q})=\sum_{s,t}(2\delta_{st}-1)\chi_{st}(\mathbf{q})\,,
\end{eqnarray*}
where $s,t\in\{\mathrm{u,d}\}$ with the spin sign convention $\mathrm{sgn}(\mathrm{u})=1$ and $\mathrm{sgn}(\mathrm{d})=-1$.

In the spin-unpolarized (paramagnetic) case of $\xi=0$, in addition to the reciprocity relations, one also has the spin symmetry relations $\chi_{\mathrm{uu}}^{(0)}(\mathbf{q})=\chi_{\mathrm{dd}}^{(0)}(\mathbf{q})=\chi_\textnormal{tot}^{(0)}(\mathbf{q})/2$, $\chi_{\mathrm{uu}}(\mathbf{q})=\chi_{\mathrm{dd}}(\mathbf{q})$ and $G_{\mathrm{uu}}(\mathbf{q})=G_{\mathrm{dd}}(\mathbf{q})$. These relations lead to~\cite{quantum_theory} 
\begin{eqnarray}
\chi_{\textnormal{tot}}^{\xi=0}(\mathbf{q})&=& \frac{\chi^{(0)}_\textnormal{tot}(\mathbf{q})}{1-\frac{4\pi}{q^2}\left[1-G_{+}(\mathbf{q})\right]\chi^{(0)}_\textnormal{tot}(\mathbf{q})}\,,\label{eq:density-density-paramagnetic}\\
\chi_{\textnormal{S}}^{\xi=0}(\mathbf{q})&=& \frac{\chi^{(0)}_\textnormal{tot}(\mathbf{q})}{1+\frac{4\pi}{q^2}G_{-}(\mathbf{q})\chi^{(0)}_\textnormal{tot}(\mathbf{q})}\,\label{eq:spin-spin-paramagnetic}\,,
\end{eqnarray}
where $G_{+}(\mathbf{q})$ and $G_{-}(\mathbf{q})$ are the symmetric and antisymmetric static LFC combinations that are defined by
\begin{eqnarray}
G_{+}(\mathbf{q})&=& \frac{G_{\mathrm{uu}}(\mathbf{q})+G_{\mathrm{ud}}(\mathbf{q})}{2}\,,\label{eq:symmetric_SLFC}\\
G_{-}(\mathbf{q})&=& \frac{G_{\mathrm{uu}}(\mathbf{q})-G_{\mathrm{ud}}(\mathbf{q})}{2}\,\label{eq:antisymmetric_SLFC}\,.
\end{eqnarray}

\begin{figure*}\centering
\includegraphics[width=0.5\textwidth]{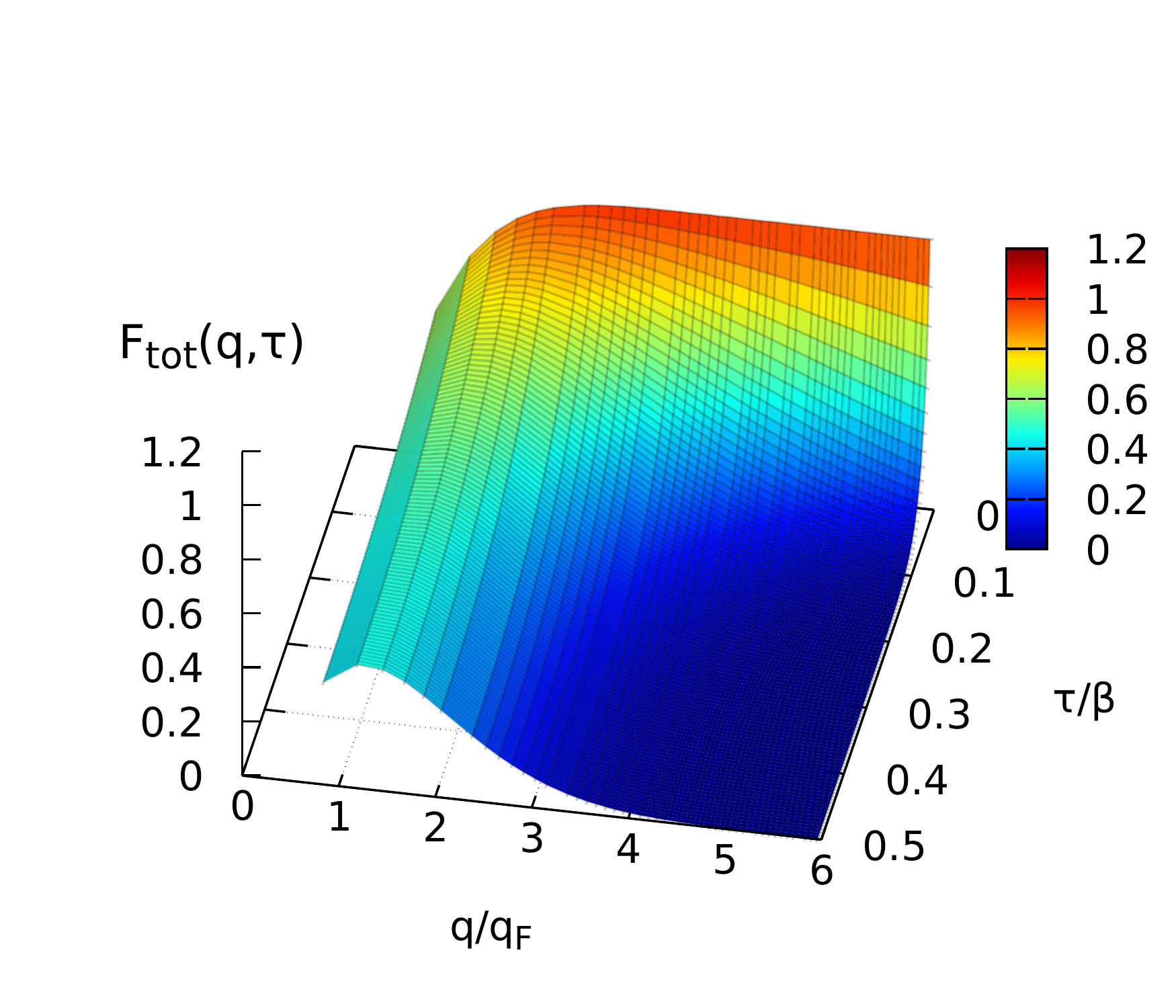}\includegraphics[width=0.5\textwidth]{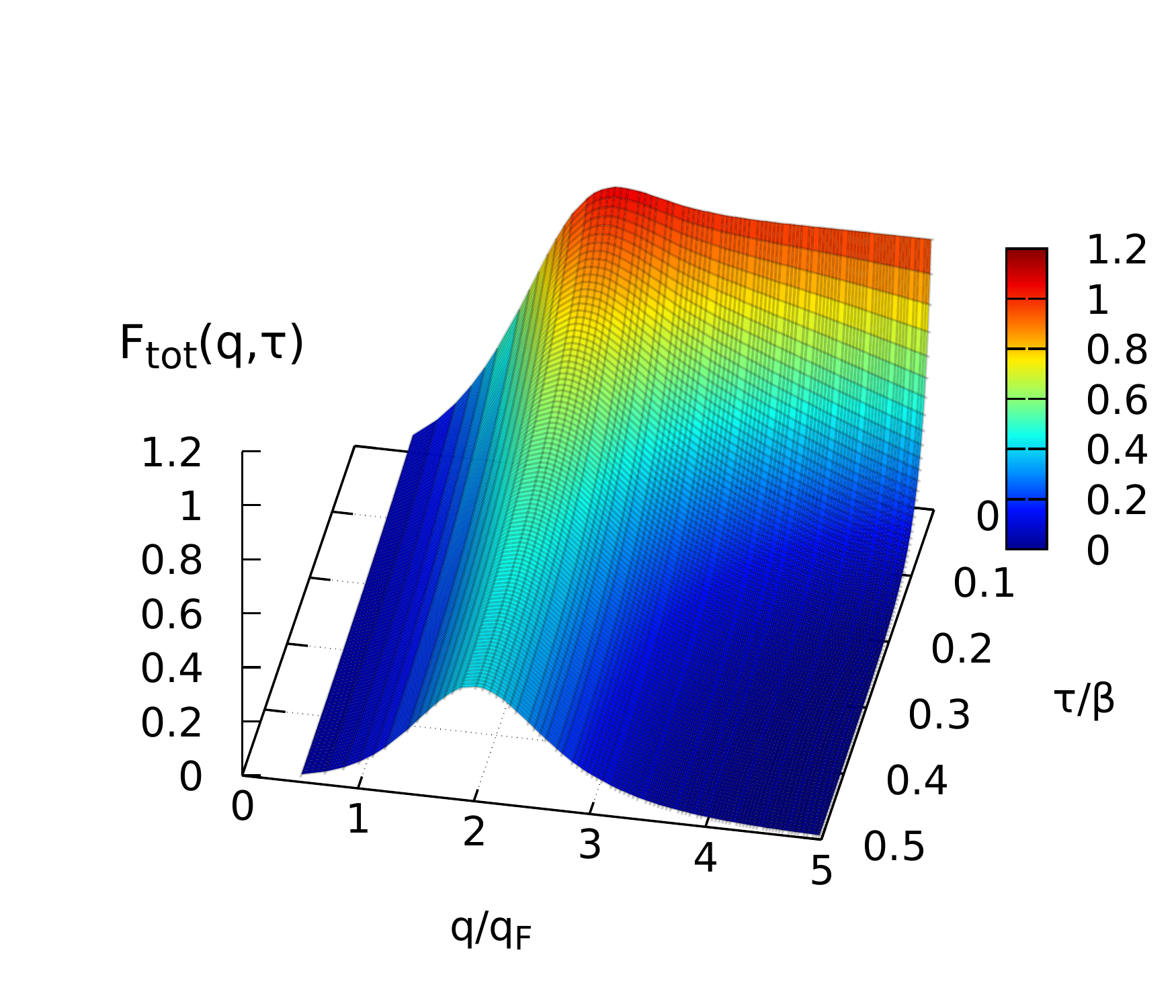}\\\vspace*{-1.8cm}\includegraphics[width=0.5\textwidth]{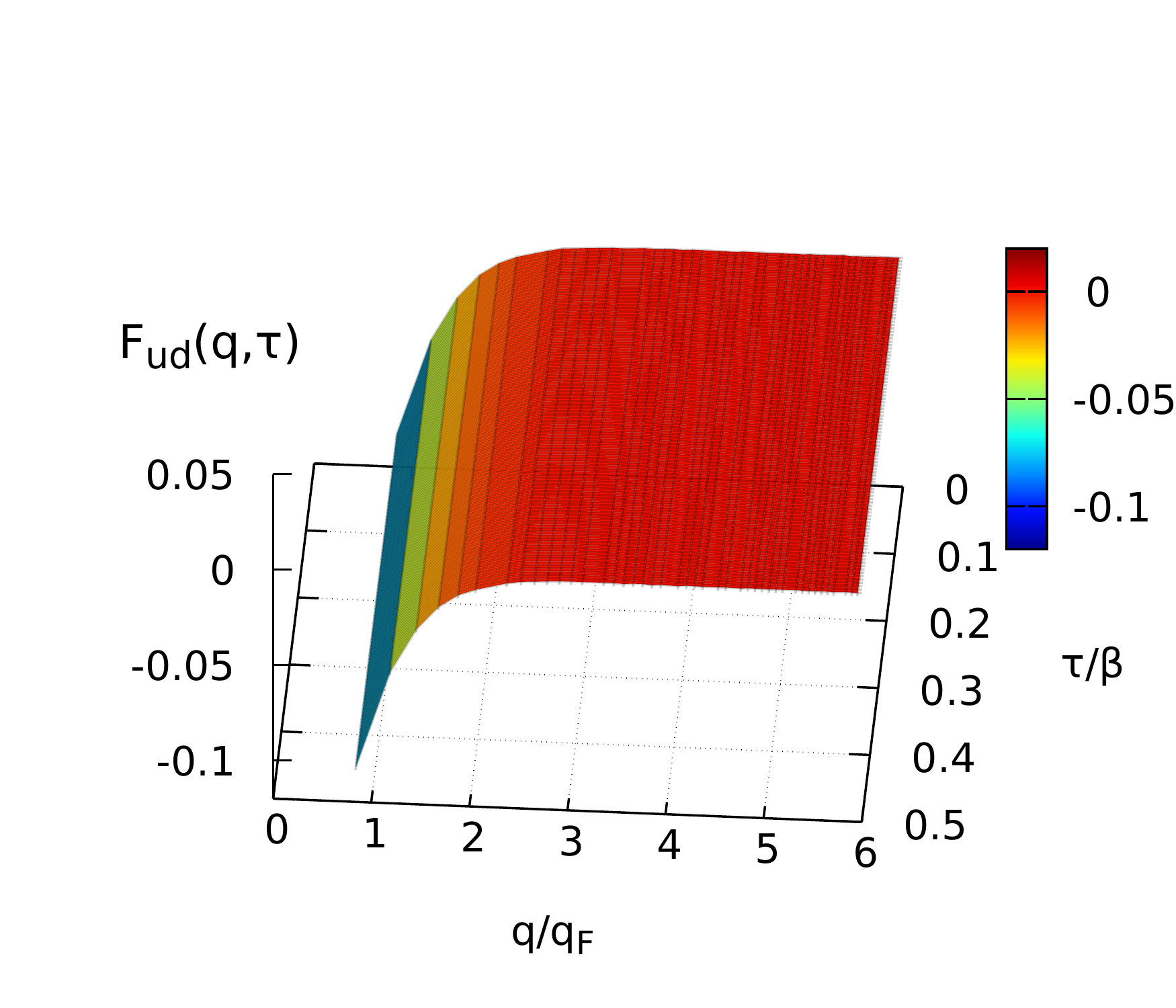}\includegraphics[width=0.5\textwidth]{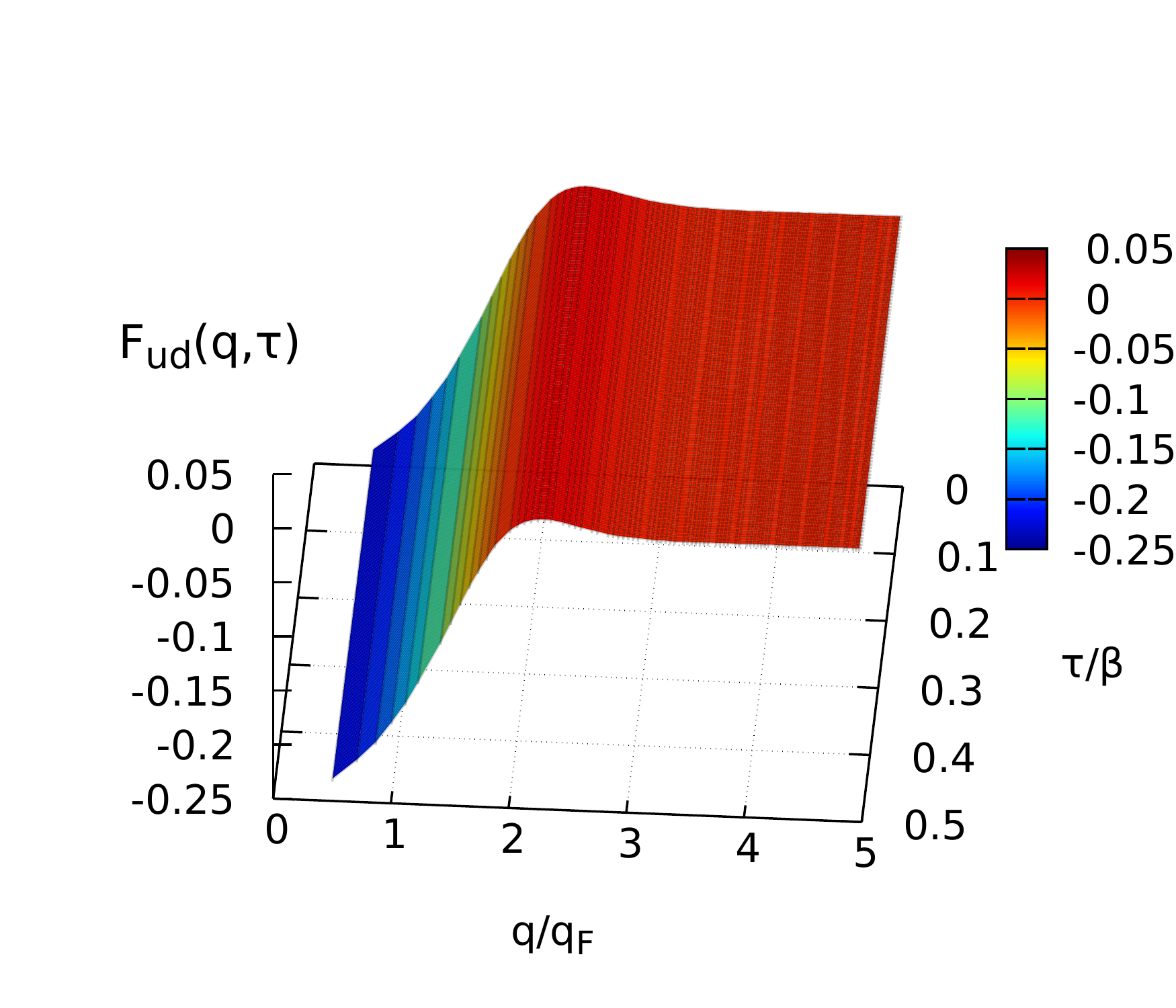}\\\vspace*{-1.57cm}\includegraphics[width=0.5\textwidth]{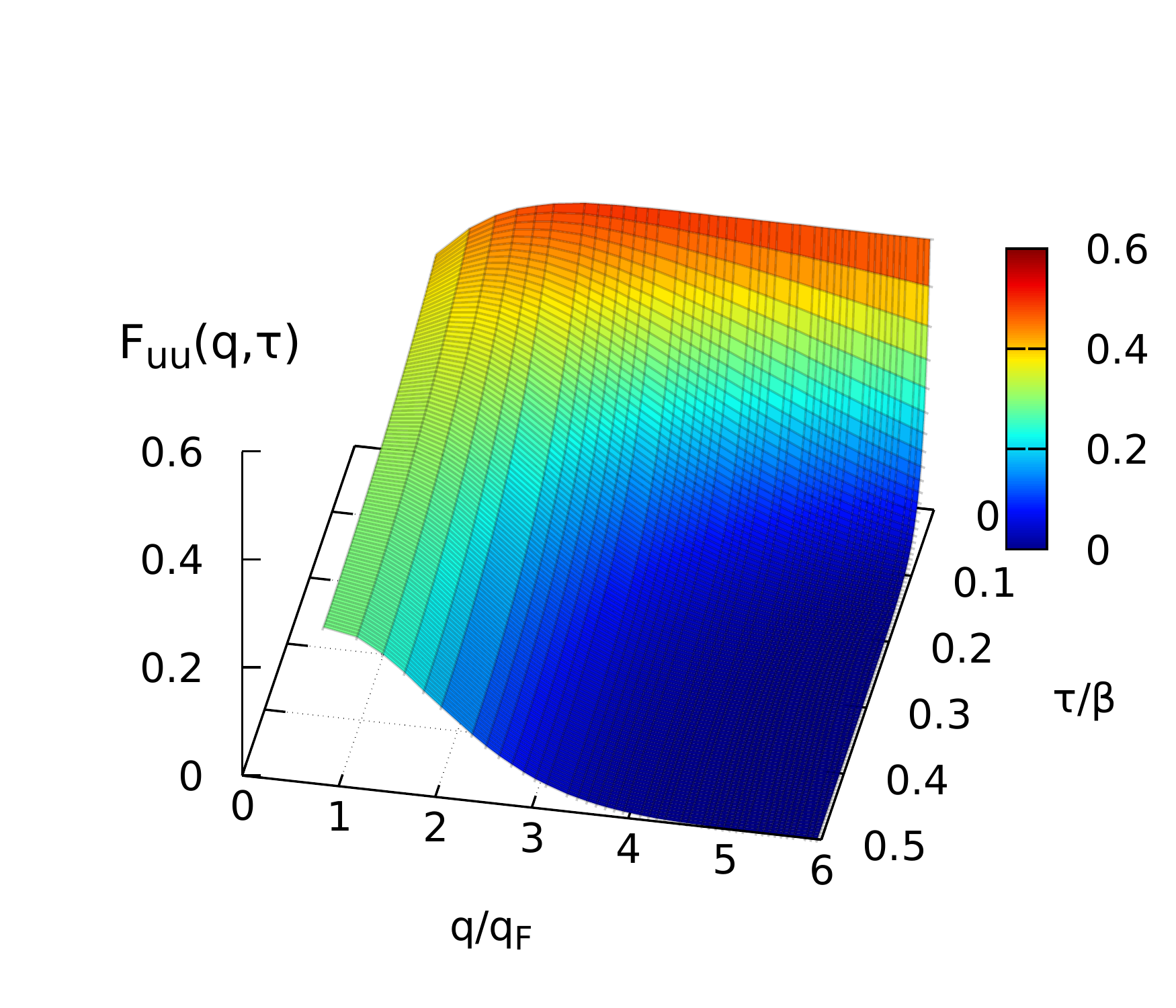}\includegraphics[width=0.5\textwidth]{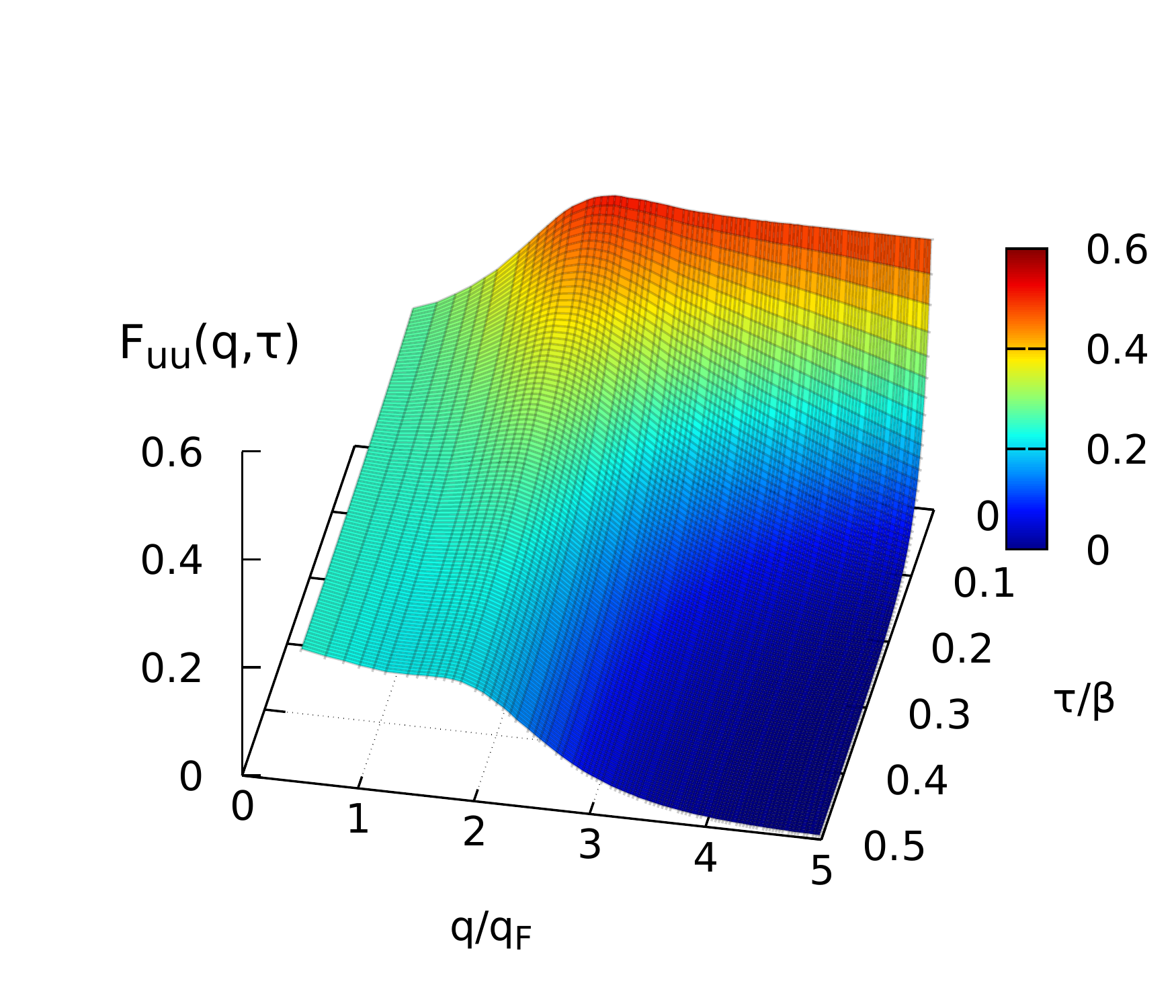}\vspace*{-0.49cm}
\caption{\label{fig:3D}
PIMC results for the spin-resolved components of the imaginary-time intermediate scattering function of the unpolarized UEG at the electronic Fermi temperature ($\theta=1$) and two quantum coupling parameters: $r_{\mathrm{s}}=2$, $N=14$ (left) and $r_{\mathrm{s}}=20$, $N=66$ (right). Top row: the total IT-ISF $F_{\mathrm{tot}}(\mathbf{q},\tau)$; center: the spin-offdiagonal IT-ISF element $F_{\mathrm{ud}}(\mathbf{q},\tau)=F_{\mathrm{du}}(\mathbf{q},\tau)$; bottom: the spin-diagonal IT-ISF element $F_{\mathrm{uu}}(\mathbf{q},\tau)=F_{\mathrm{dd}}(\mathbf{q},\tau)$.
}
\end{figure*}

\section{Results\label{sec:results}}

\begin{figure*}\centering
\includegraphics[width=0.475\textwidth]{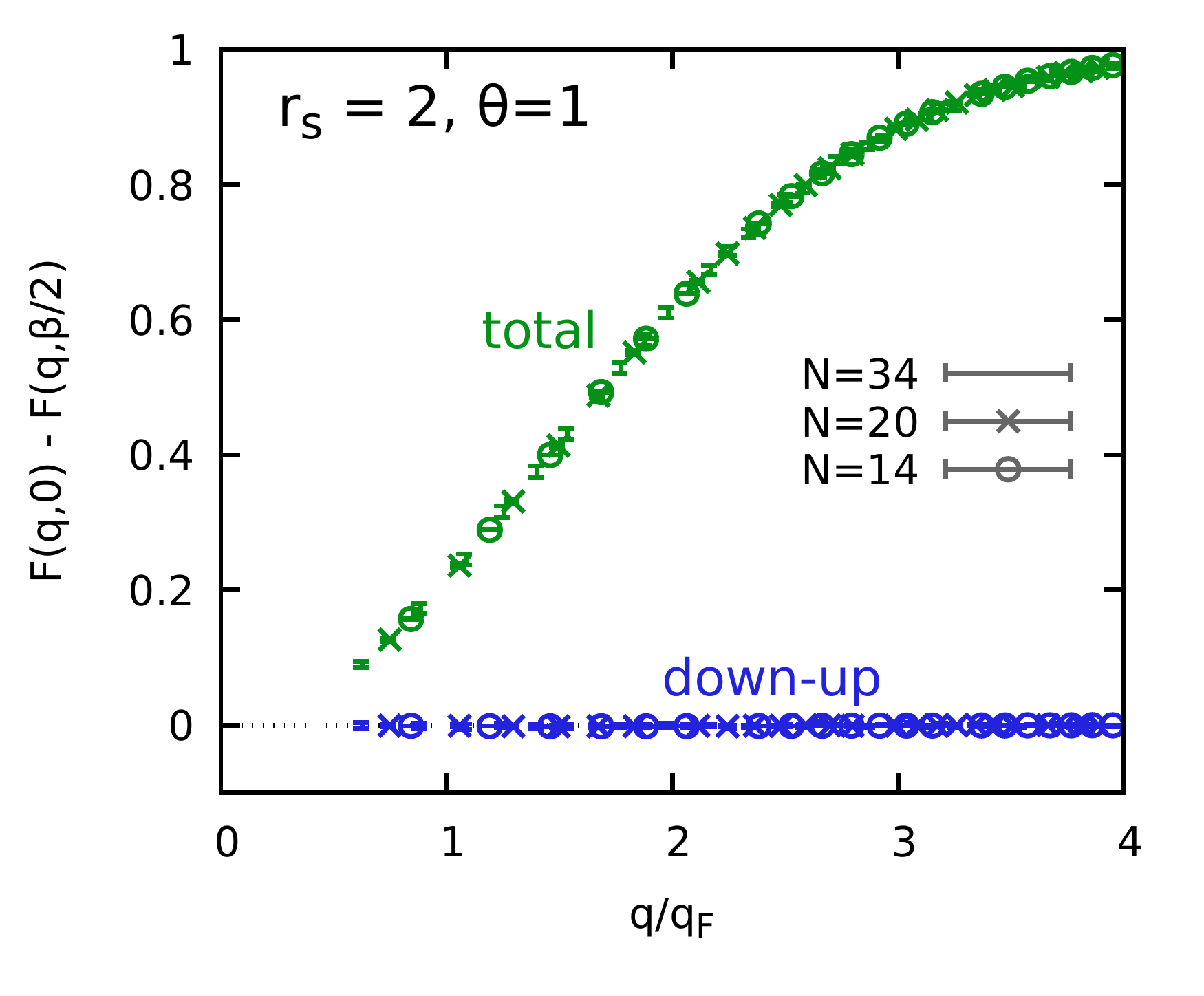}\includegraphics[width=0.475\textwidth]{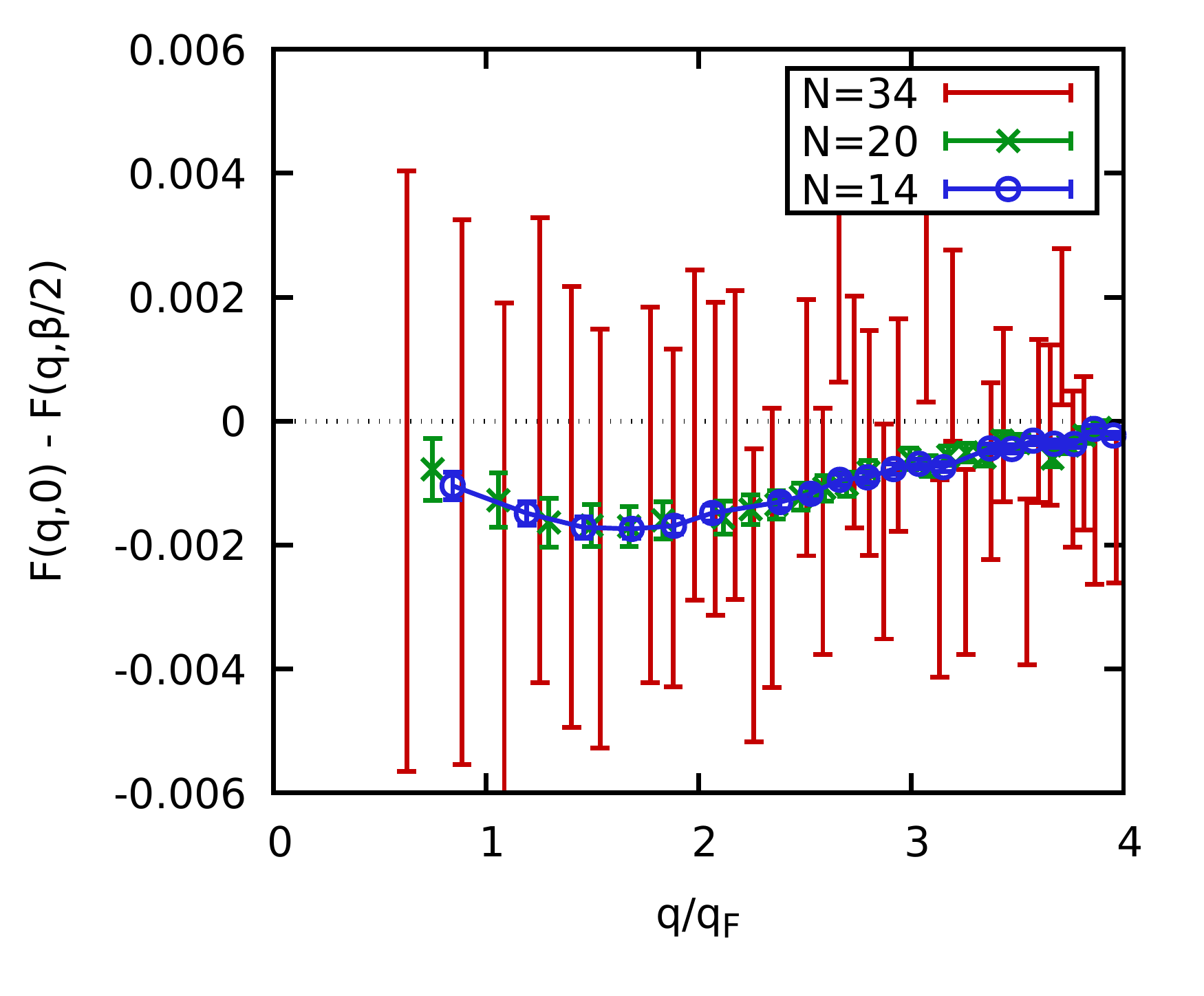}
\caption{\label{fig:tau_dependence}
Decay along the $\tau$-direction of the imaginary-time intermediate scattering function quantified by $\Delta F_\tau(q)=F(q,0) - F(q,\beta/2)$ at $r_s=2$ and $\theta=1$. Left: PIMC results for different $N$ for $F_\textnormal{tot}$ (green) and its spin-offdiagonal component $F_{\mathrm{du}}$ (blue). Right: Magnified segment around $F_{\mathrm{du}}$ for different $N$.
}
\end{figure*} 

All the PIMC simulations are carried out using an implementation of the \emph{extended ensemble} approach introduced in Ref.~\cite{Dornheim_PRB_nk_2021}, which can be viewed as a canonical adaption of the seminal \emph{worm algorithm} by Boninsegni \emph{et al.}~\cite{boninsegni1,boninsegni2}. Since detailed introductions to PIMC have been presented elsewhere~\cite{cep,boninsegni1}, here we only summarize the main technical aspects of our investigation. In particular, we do not impose any restrictions on the nodal structure of the thermal density matrix~\cite{Ceperley1991}. Therefore, our simulations are exact within the given Monte Carlo error bars, but they are afflicted by the notorious fermion sign problem~\cite{troyer,dornheim_sign_problem}. The latter manifests as an exponential increase in the required compute time with respect to system parameters like the number of electrons $N$, or the inverse temperature $\beta=1/k_\textnormal{B}T$. Indeed, the sign problem constitutes the central practical bottleneck in our simulations and limits their application to the temperature range $0.75\leq\theta$. For completeness, we note that we employ the primitive high-temperature factorization $e^{-\beta\hat{H}/P}\approx e^{-\beta\hat{K}/P}e^{-\beta\hat{V}/P}$ (with $\hat{K}$ and $\hat{V}$ being the kinetic and potential contributions to the full Hamiltonian Eq.~(\ref{eq:Hamiltonian}), respectively), and the convergence with the number of factors $P$ has been carefully checked. Higher-order decompositions of $e^{-\beta\hat{H}/P}$ have been studied in the literature~\cite{sakkos_JCP_2009,Zillich_JCP_2010}, but are not required for the conditions studied in the present work. All PIMC results are freely available online~\cite{repo} and can be used as a benchmark for approximations and new developments.

\subsection{Spin-resolved imaginary-time intermediate scattering function\label{sec:ITCF_results}}

Let us start our investigation of the spin-resolved density response of the warm dense electron gas by considering the imaginary-time density--density correlation function $F(\mathbf{q},\tau)$, which constitutes the basis for most results obtained in this work. In Fig.\ref{fig:3D}, we plot its different components at the electronic Fermi temperature $\theta=1$ for $r_{\mathrm{s}}=2$ (left column) and for $r_{\mathrm{s}}=20$ (right column). More specifically, $r_{\mathrm{s}}=2$ corresponds to a typical metallic density that can be realized experimentally for example in aluminum~\cite{Sperling_PRL_2015,aluminum1}. 
The top left panel of Fig.~\ref{fig:3D} concerns PIMC results for the total IT-ISF $F(\mathbf{q},\tau)$, cf.~Eq.~(\ref{eq:define_F}). First, it should be noted that all IT-ISF components are symmetric around $\tau=\beta/2$, $F(\mathbf{q},\tau) = F(\mathbf{q},\beta-\tau)$, thus it is sufficient to analyse this quantity in the halved domain of $0\leq\tau\leq\beta/2$. In addition, it converges towards the static structure factor $S(q)$ in the limit of $\tau\to0$, see Eq.~(\ref{eq:S_st}) above. An additional feature of the full IT-ISF depicted in the top panel concerns the monotonic decay along the $\tau$-direction, which is a direct consequence of the kernel $e^{-\tau\omega}$ of the Laplace transform Eq.~(\ref{eq:Laplace}) discussed in Sec.~\ref{sec:ITCF}. In fact, it can be shown that the steep decay of $F(\mathbf{q},\tau)$ along the imaginary time for large $q$ directly reflects the single-particle limit of the dispersion relation, which scales as $\omega(q)\sim q^2$. 

We next consider the spin-resolved contributions to the IT-ISF, starting with the spin-offdiagonal component $F_{\mathrm{ud}}(\mathbf{q},\tau)=F_{\mathrm{du}}(\mathbf{q},\tau)$ shown in the center-left panel of Fig.~\ref{fig:3D}. Evidently, there are hardly any imaginary-time dependent correlations between particles of different species, and it holds $F_{\mathrm{ud}}(\mathbf{q},\tau)\approx S_{\mathrm{ud}}(\mathbf{q})$ in good approximation. This reflects the near absence of the imaginary-time diffusion processes in this case. In other words, there certainly exist substantial correlations between particles of different spin-orientations, but they almost do not depend on the $\tau$-argument in the spin-resolved single-particle density operator $\hat{n}_s(\mathbf{q},\tau)$ defined in Eq.~(\ref{eq:spin_resolved_density}).

To further investigate this remarkable observation, we rigorously quantify the $\tau$-dependence of both $F_\textnormal{tot}(\mathbf{q},\tau)$ (green data points) and $F_\mathrm{ud}(\mathbf{q},\tau)$ (blue data points) in Fig.~\ref{fig:tau_dependence}. Specifically, we consider the overall decay along $\tau$, which we quantify as $\Delta F_\tau(q)=F(q,0) - F(q,\beta/2)$. The bars, crosses, and circles depict our PIMC results for $N=34$, $N=20$, and $N=14$ unpolarized electrons. Clearly, no dependence on the system size can be resolved within the given error bars, which is consistent to previous investigations of the IT-ISF of the UEG at similar parameters~\cite{Dornheim_PRE_2020}. The green points again show the substantial $\tau$-dependence of the total function $F_\textnormal{tot}(\mathbf{q},\tau)$ for all wave numbers, and the decay increases with $q$. In the large-$q$ limit, it holds $F(\mathbf{q},0)=S(\mathbf{q})=1$ and $F(\mathbf{q},\beta/2)=0$, which means that $\lim_{q\to\infty}\Delta F_\tau(q)=1$. On the other hand, no $\tau$-dependence can be seen in the spin-offdiagonal IT-ISF on the depicted scale. To decisively resolve the possible $\tau$-dependence of this function, we show a magnified segment around the up-down component of $\Delta F_\tau$ in the right panel of Fig.~\ref{fig:tau_dependence}, with the different symbols corresponding to different numbers of electrons $N$. On this scale, we can clearly resolve a small, though statistically significant dependence on $\tau$ independent of $N$; we note that the comparably large error bars for $N=34$ are a direct consequence of the exponential increase in compute time with the system size due to the fermion sign problem~\cite{dornheim_sign_problem}. In other words, imaginary-time diffusion between electrons of different species does play a role, but orders of magnitudes smaller than the effects of self-diffusion in $F_\textnormal{tot}(\mathbf{q},\tau)$.

\begin{figure*}\centering
\includegraphics[width=0.475\textwidth]{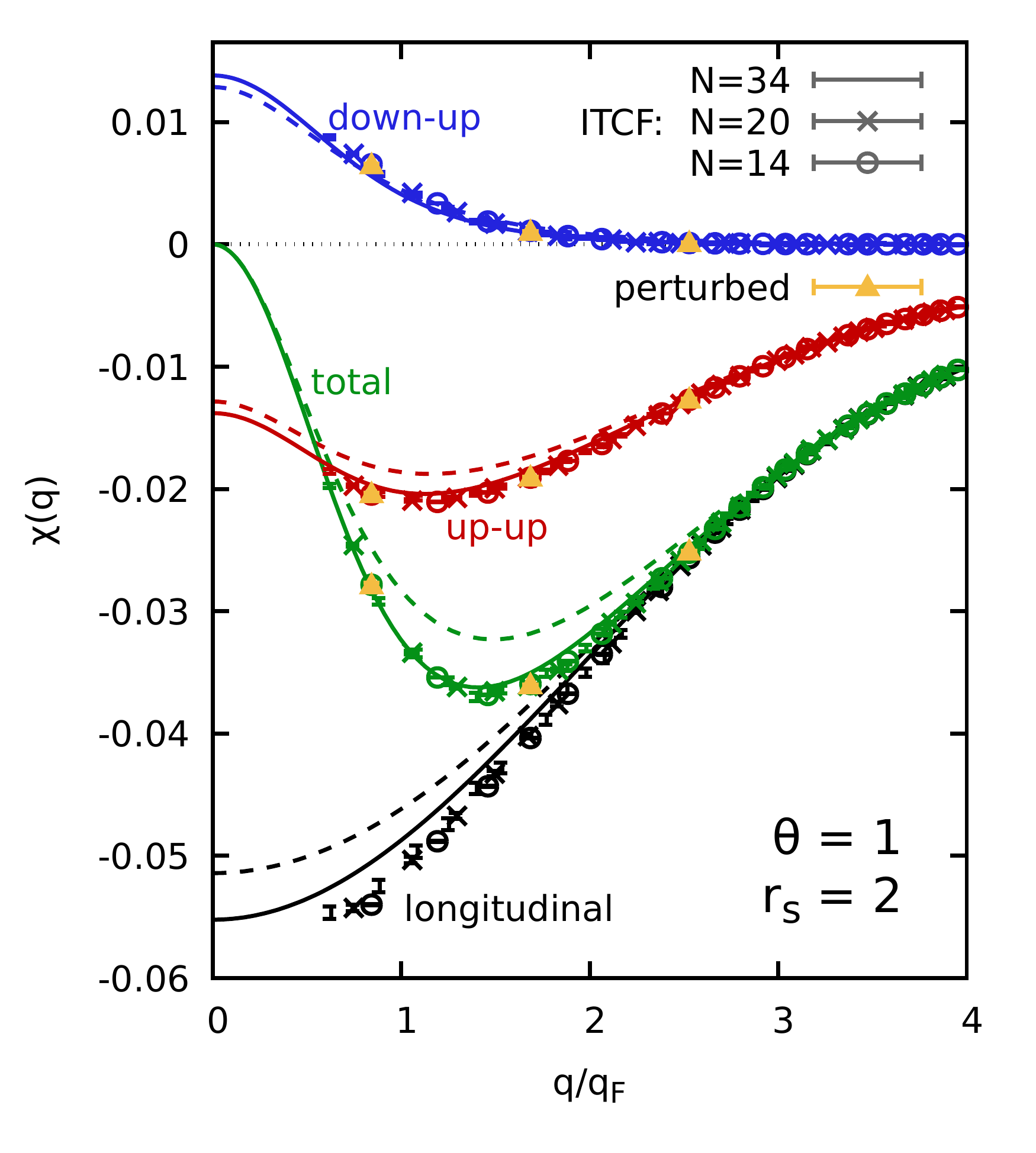}\includegraphics[width=0.475\textwidth]{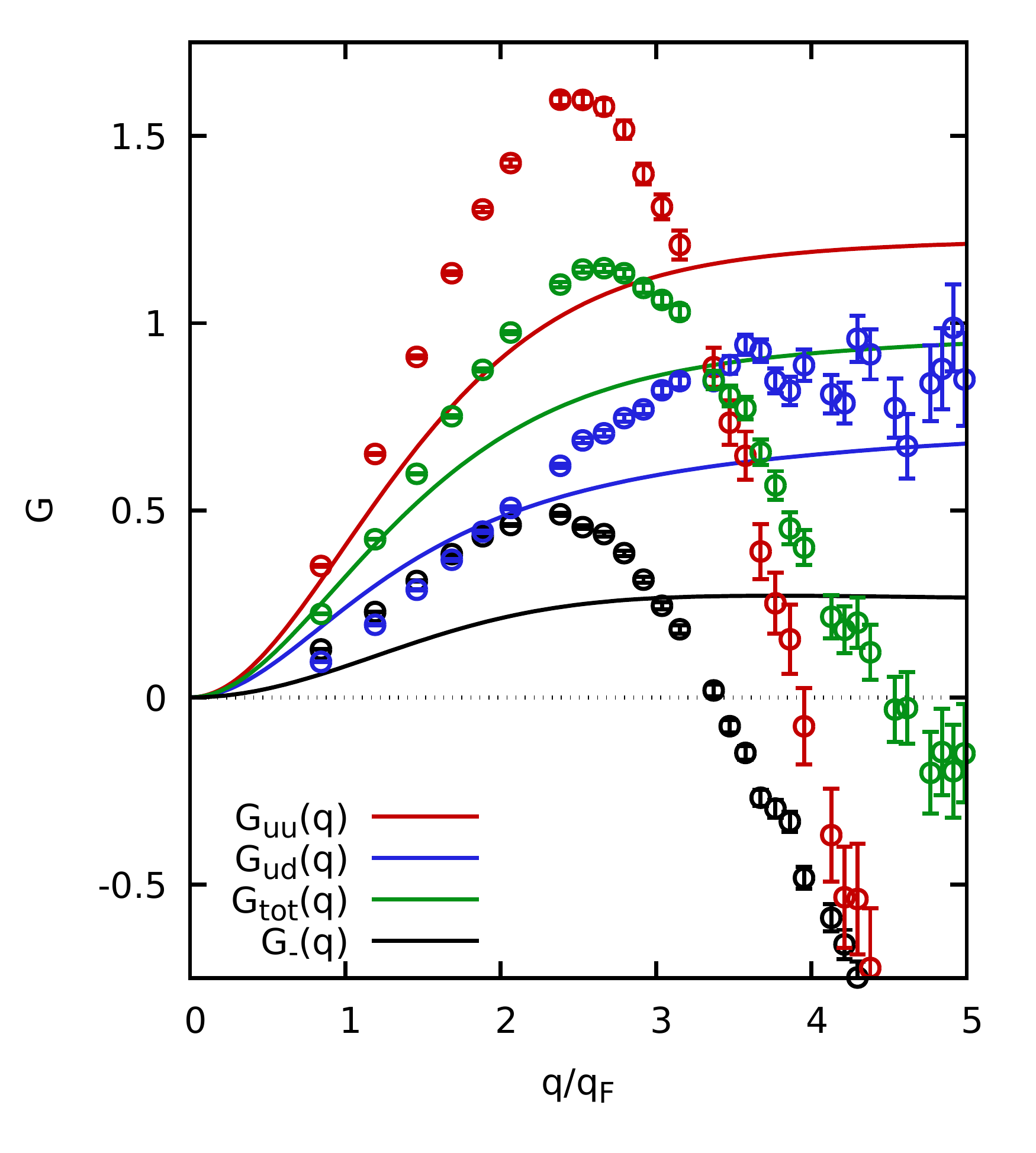}
\caption{\label{fig:q_dependence}
Left: The spin-resolved components of the static density response of the unpolarized UEG at $r_{\mathrm{s}}=2$ and $\theta=1$. The bars, crosses, and circles correspond to PIMC results for $N=34$, $N=20$, and $N=14$, respectively, while the dashed and solid lines have been obtained within the finite-temperature RPA and the STLS scheme~\cite{stls,stls2}. The yellow triangles result from the fit to the density response of the harmonically perturbed UEG, cf.~Fig.~\ref{fig:A_dependence}. Right: The corresponding results for the spin resolved LFCs, see Eq.~(\ref{eq:LFC}).}
\end{figure*} 

\begin{figure*}\centering
\includegraphics[width=0.475\textwidth]{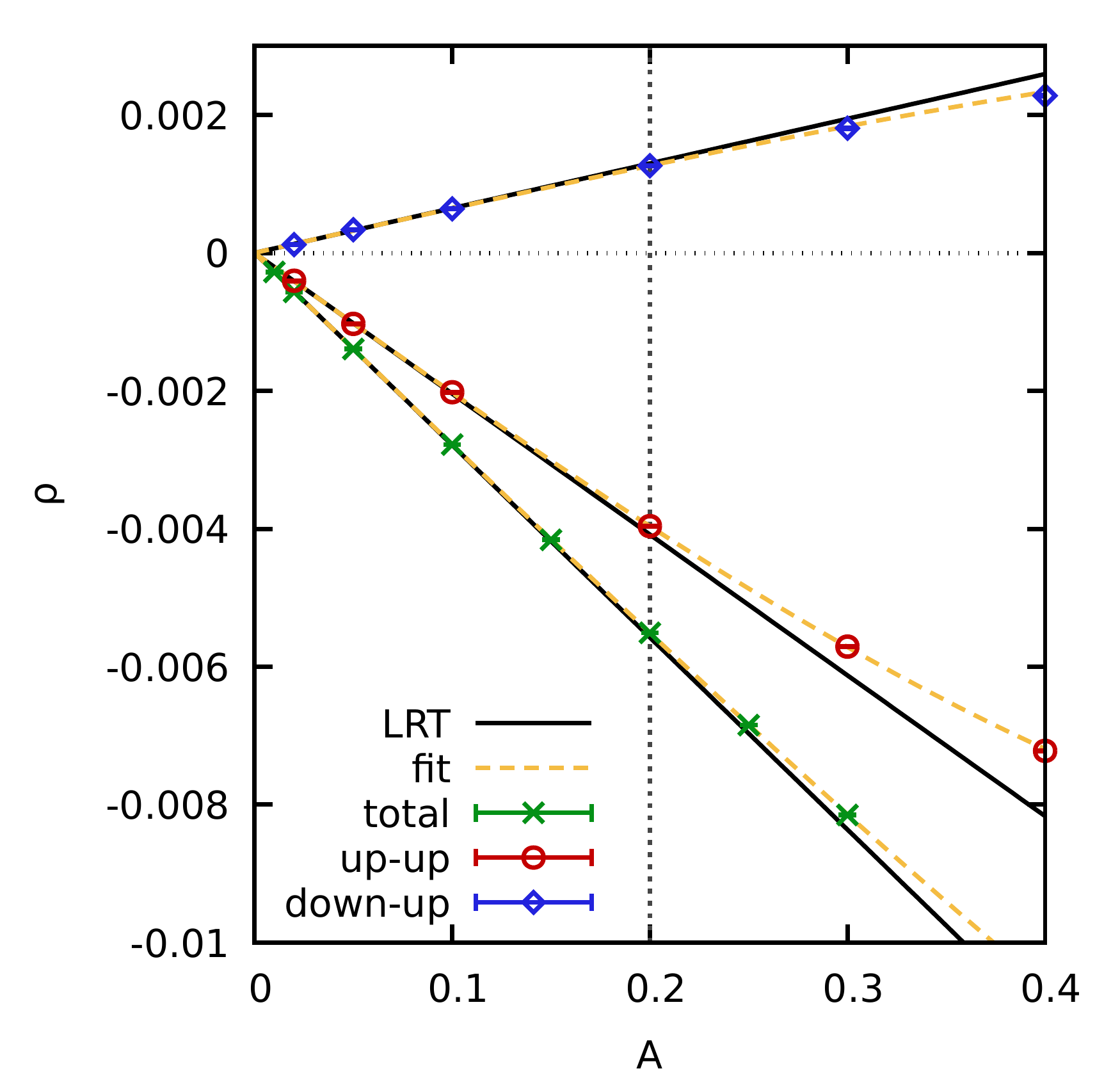}\includegraphics[width=0.475\textwidth]{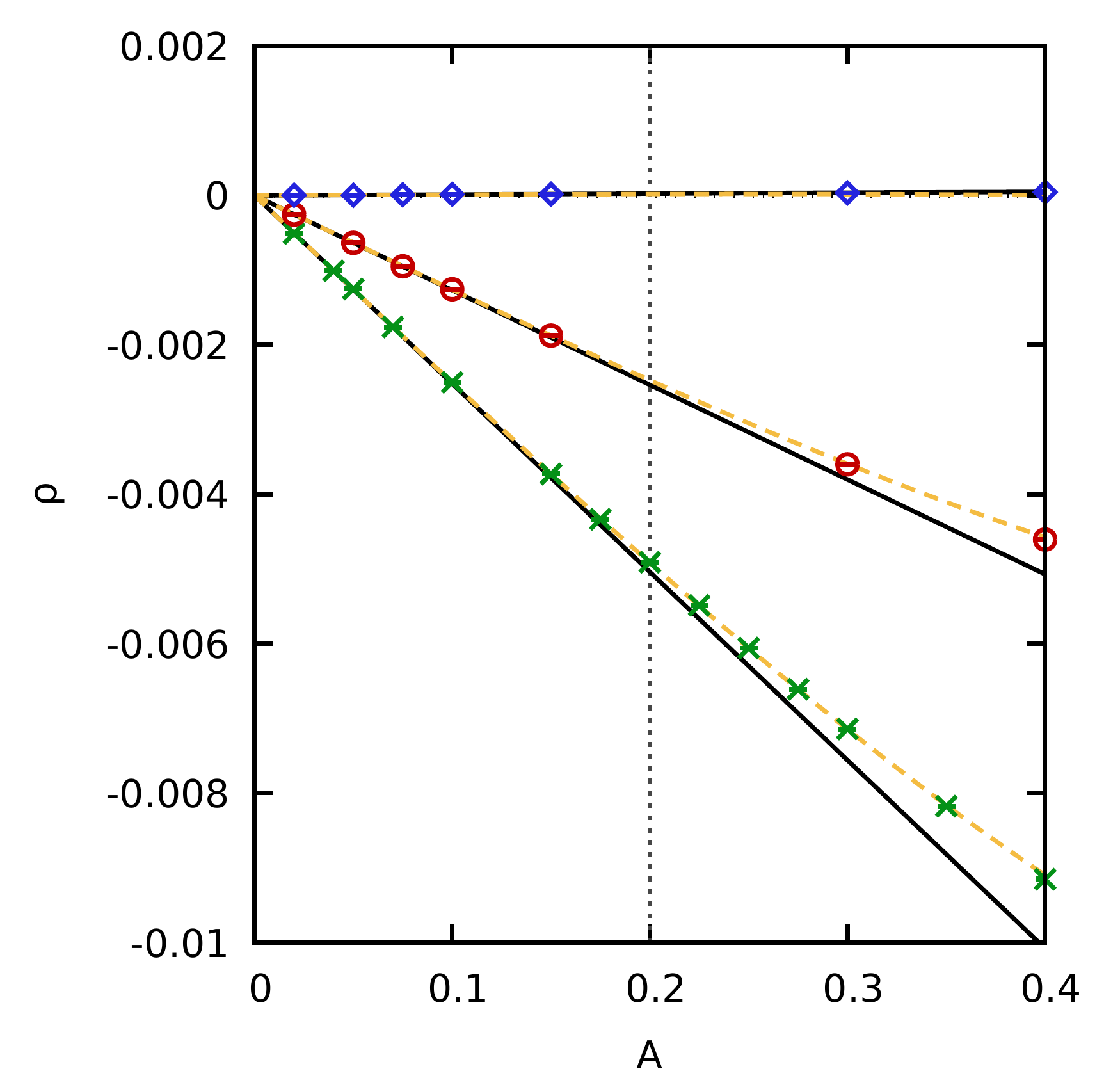}
\caption{\label{fig:A_dependence}
Plot of the spin-resolved density response versus the perturbation amplitude for the same conditions as in Fig.~\ref{fig:q_dependence}, after only the spin-up electrons have been perturbed [i.e., $A_{\mathrm{u}}>0$ and $A_{\mathrm{d}}=0$ in Eq.~(\ref{eq:Hamiltonian})], but the response of both spin components has been measured. Blue diamonds: the spin-offdiagonal response $\rho_{\mathrm{d}}(\mathbf{q},A_{\mathrm{u}})$; red circles: the spin-diagonal response $\rho_{\mathrm{u}}(\mathbf{q},A_{\mathrm{u}})$; green crosses: the total density response of the unpolarized UEG to a response acting on both spin components, adopted from Refs.~\cite{Dornheim_PRL_2020,Dornheim_PRR_2021}. Left: $\mathbf{q}=2\pi L^{-1}(1,0,0)^T$ ($q\approx0.84q_\textnormal{F}$); right: $\mathbf{q}=2\pi L^{-1}(3,0,0)^T$ ($q\approx2.52q_\textnormal{F}$). The solid black line corresponds to the prediction of linear-response theory, and the dashed yellow line corresponds to the cubic fit according to Eq.~(\ref{eq:fit}).
}
\end{figure*} 

The IT-ISF that measures the imaginary-time density correlations between electrons of the same spin orientation $F_{\mathrm{uu}}(\mathbf{q},\tau)=F_{\mathrm{dd}}(\mathbf{q},\tau)$ is plotted at the bottom of Fig.~\ref{fig:3D} and qualitatively resembles the full IT-ISF $F(\mathbf{q},\tau)$ shown at the top. 

We shall conclude this analysis by considering the substantially decreased density, $r_{\mathrm{s}}=20$, that is plotted in the right column of Fig.~\ref{fig:3D}. This corresponds to a substantially more strongly coupled system (the classical coupling parameter is now $\Gamma\approx0.54r_{\mathrm{s}}/\theta\approx10$~\cite{Ott2018}) that belongs to the electron liquid regime~\cite{dornheim_electron_liquid,quantum_theory,Tolias_JCP_2021}. Since the basic behaviour of all IT-ISFs is the same compared to $r_s=2$, we restrict ourselves to a brief discussion of the main differences due to the increased coupling strength. First and foremost, we find a more pronounced structure in all IT-ISFs, with a distinct, correlation-induced peak located around twice the Fermi wave number, $q\sim2q_\textnormal{F}$. The physical origin of this effect can be traced back to the spontaneous excitation of aligned pairs of electrons~\cite{Dornheim_Nature_2022}, which have a reduced interaction energy in the presence of the UEG~\cite{Dornheim_Force_2022}.
This peak exists throughout the entire $\tau$-domain, and in all individual components of $F(\mathbf{q},\tau)$.
In addition, we point out that the observed structure of the IT-ISF at $r_s=20$ shown in Fig.~\ref{fig:3D} leads to a roton-like feature in the dynamic structure factor $S(\mathbf{q},\omega)$, which is discussed in detail in the recent Ref.~\cite{Dornheim_Nature_2022}.

\subsection{Spin-resolved density response at metallic densities\label{sec:spin}}

We proceed with the central task of the present investigation, which is the accurate estimation of the spin-resolved density response $\chi_{st}(q)$ to an external static harmonic perturbation. More specifically, the latter may affect either both, or only a single electron species. In the left panel of Fig.~\ref{fig:q_dependence}, we show our new PIMC results that have been obtained by integrating the respective IT-ISF along the $\tau$-direction at $r_s=2$ and $\theta=1$. In particular, the different symbols depict our PIMC data for $N=34$ (bars), $N=20$ (crosses), and $N=14$ (circles) unpolarized electrons, and the red, blue, and green colours correspond to the up-up component, the down-up component, and the full density response function, respectively. First and foremost, we find no significant dependence of our results on the system size for as few as $N=14$ particles. This is consistent with previous studies of wavenumber resolved properties of the UEG~\cite{Chiesa_PRL_2006,Drummond_PRB_2008,dornheim_prl,Holzmann_PRB_2016,dornheim_cpp,Dornheim_PRE_2020,Dornheim_JCP_2021}. Physically, the response of the spin-down electrons to a perturbation of the spin-up electrons, $\chi_{\mathrm{du}}(\mathbf{q})$, has the opposite sign of the spin-diagonal response function $\chi_{\mathrm{uu}}(\mathbf{q})$, and the two exactly cancel in the limit of $q\to0$. The former can be understood as follows: the spin-up electrons will, on average, move to the minima of the applied external potential. Therefore, their response has a negative sign, $\chi_{\mathrm{uu}}(\mathbf{q})<0$. This, in turn, induces the unperturbed spin-down electrons to move away from the other species, and to occupy the now less populated space where the external field on the spin-up electrons is large. Consequently, the corresponding response function has a positive sign. The latter can be understood in view of the exact long wavelength behavior of the UEG total static density response, that is given by~\cite{kugler_bounds}
\begin{eqnarray}\label{eq:perfect_screening}
\lim_{q\to0}\chi_\textnormal{tot}(\mathbf{q}) = - \frac{q^2}{4\pi}\ .
\end{eqnarray}
which is a direct consequence of the perfect screening in the UEG. A second notable difference between $\chi_{\mathrm{uu}}(\mathbf{q})$ and $\chi_{\mathrm{du}}(\mathbf{q})$ is given by the fact that $\chi_{\mathrm{du}}(\mathbf{q})$ already decayed close to zero for $q\gtrsim2q_\textnormal{F}$, whereas $\chi_{\mathrm{uu}}(\mathbf{q})$ remains significant for substantially larger values of $q$. The physical origin of this effect can be understood by considering the role of the wavelength $\lambda=2\pi/q$ on both response functions; this is explained in detail in the discussion of Fig.~\ref{fig:spatial} below. In short, the spin-up electrons are induced to react to the external potential even on small length scales, i.e., at large $q$. In contrast, the unperturbed spin-down electrons are increasingly less affected by such small displacements of the spin-up electrons. As a consequence, $\chi_{\mathrm{du}}(\mathbf{q})$ converges to zero when $\lambda\lesssim r_{\mathrm{s}}$, which also implies that the total density density response and the longitudinal spin-spin response become nearly indistinguishable for $q\gtrsim2q_\textnormal{F}$. Concerning the $\chi_{\mathrm{S}}(\mathbf{q})$  spin response function, it is important to point out that its long wavelength limit is not zero but negative, in accordance with the spin susceptibility sum rule and the positive static paramagnetic susceptibility~\cite{SingwiTosi_Review}. Notice that the non-positivity of $\chi_{\mathrm{S}}(\mathbf{q})$ is guaranteed in nearly the entire wavenumber range by the earlier discussed $\chi_{\mathrm{du}}(\mathbf{q})>0$ and $\chi_{\mathrm{uu}}(\mathbf{q})<0$, since $\chi_{\mathrm{S}}(\mathbf{q})=2[\chi_{\mathrm{uu}}(\mathbf{q})-\chi_{\mathrm{du}}(\mathbf{q})]$ in the unpolarized UEG.

The dashed lines show the widespread random phase approximation (RPA), which describes the density response of the UEG to an external perturbation on the mean-field level and whose two-component version assumes that all the spin-resolved LFCs are identically zero. Evidently, it is in good qualitative agreement with the PIMC data for all depicted components at these parameters. Indeed, the RPA is well known to exactly describe the total density response in the limit of large wavelengths, see Eq.~(\ref{eq:perfect_screening}). The most pronounced systematic errors in the RPA appear around $q\sim1.5q_\textnormal{F}$, where the total density response attains a maximum. This follows from the addition of the deviations in both components $\chi_{\mathrm{uu}}(\mathbf{q})$ and $\chi_{\mathrm{ud}}(\mathbf{q})$ as the systematic errors have the same sign for intermediate $q$. From a physical perspective, the deficiency can be traced back to the spontaneous alignment of electron pairs in the UEG, which sensitively depends on the accurate description of the effective interaction potential~\cite{Dornheim_Nature_2022,Dornheim_Force_2022}.

Interestingly, the RPA exhibits systematic errors for $q\to0$ in the individual components, which only cancel in the total response function $\chi_{\mathrm{tot}}(\mathbf{q})$. This can be seen particularly well by comparing them to the solid curves, which have been obtained on the basis of the approximate static LFC that was originally suggested by Singwi, Tosi, Land, and Sj\"olander~\cite{stls_original} (STLS); see Refs.~\cite{IIT,stls,stls2} for the extension of this idea to finite temperatures. Evidently, the STLS scheme also attains the correct limit of the total response function, i.e., Eq.~(\ref{eq:perfect_screening}). Yet, it predicts a different $q\to0$ limit for the individual components spin resolved compared to RPA. In addition, we observe that the STLS scheme constitutes a substantial improvement over the RPA over the entire $q$-range, and the remaining deviations to the PIMC data are small. Finally, we emphasize that the favorable cancellation of errors in the RPA and STLS predictions for the long-wavelength total density response function, unavoidably turns into an unfavorable augmentation of errors in the RPA and STLS predictions for the long wavelength spin response function. Concerning the $\chi_{\mathrm{S}}(\mathbf{q})$ response, the STLS scheme still constitutes a substantial improvement over the RPA within the entire $q$-range, but its $q\to0$ deviations to the PIMC data are rather substantial.

\begin{figure*}\centering
\includegraphics[width=0.475\textwidth]{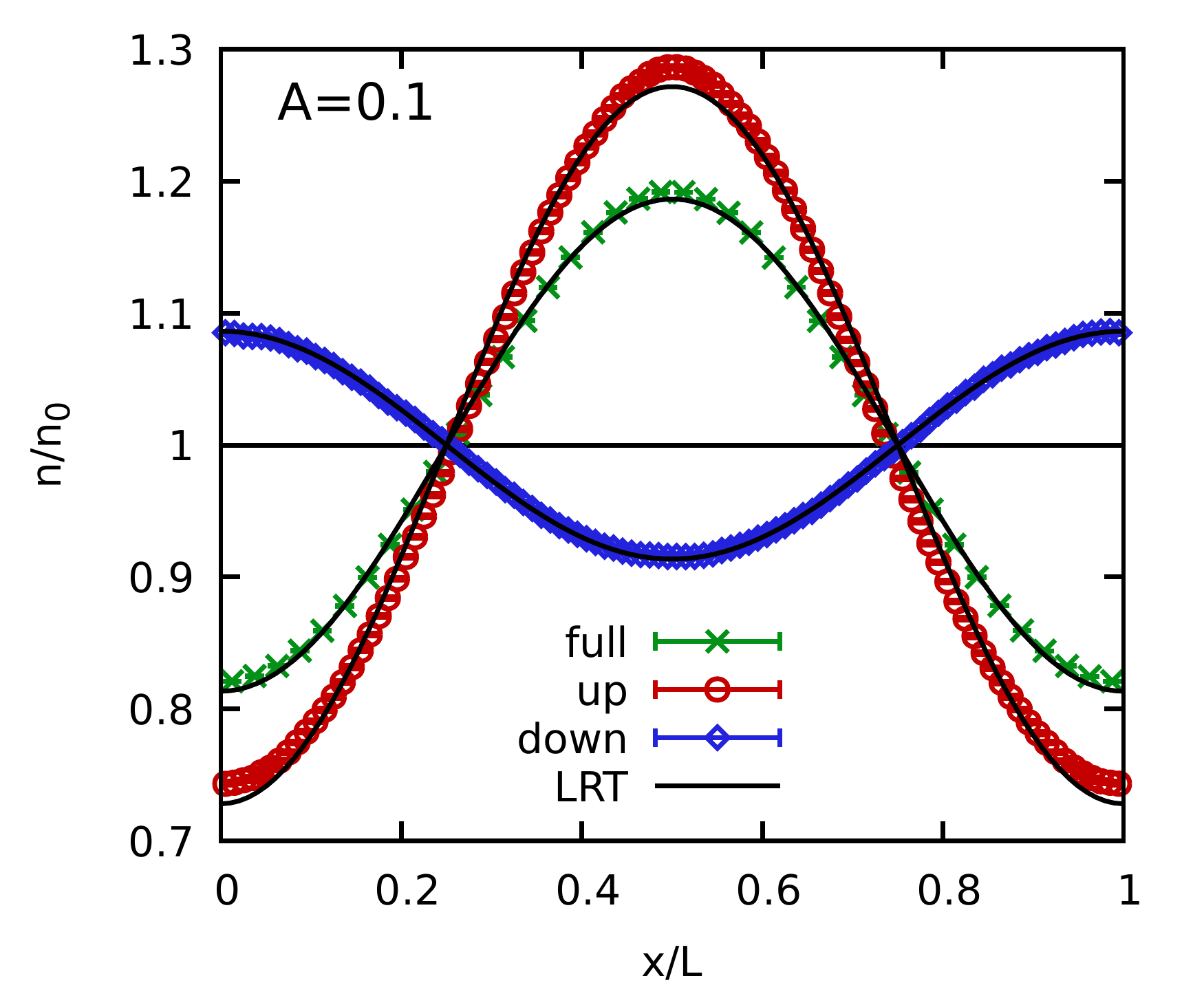}\includegraphics[width=0.475\textwidth]{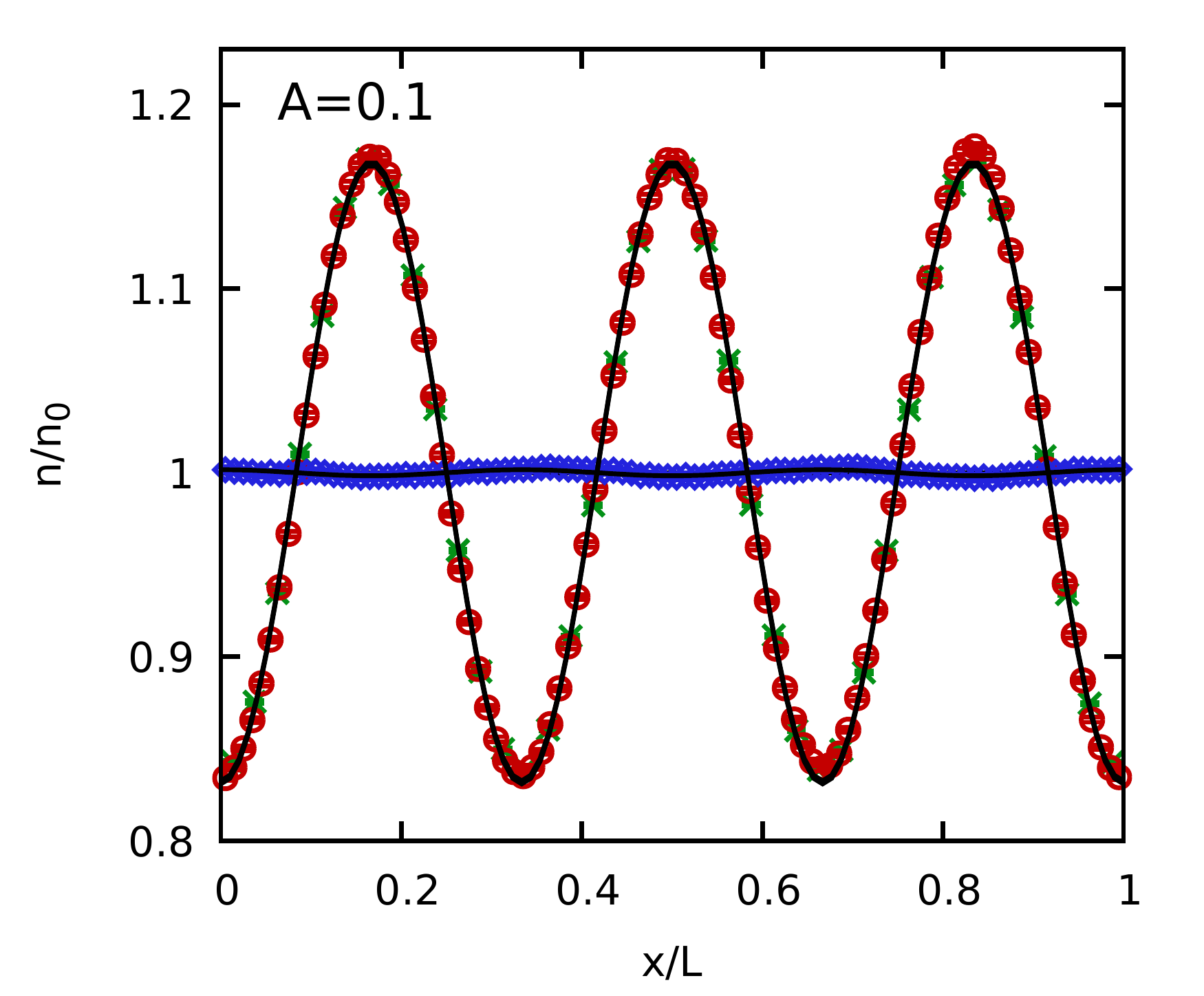}
\caption{\label{fig:Density}
Plot of the relative density profile (with respect to the average density) in coordinate space for the same conditions as in Fig.~\ref{fig:A_dependence} and the perturbation amplitude $A=0.1$. Red circles (blue diamonds): the density response of the spin-up (spin-down) electrons to a perturbation with $A_{\mathrm{u}}=0.1$ and with $A_{\mathrm{d}}=0.1$. Green crosses: the density response of both spin-components to a perturbation with $A_{\mathrm{u}}=A_{\mathrm{d}}=0.1$, adopted from Ref.~\cite{Dornheim_PRR_2021}.
Left: $\mathbf{q}=2\pi L^{-1}(1,0,0)^T$ ($q\approx0.84q_\textnormal{F}$); right: $\mathbf{q}=2\pi L^{-1}(3,0,0)^T$ ($q\approx2.52q_\textnormal{F}$).
}
\end{figure*}

We shall postpone the discussion of the yellow triangles for now, and instead consider the spin-resolved static LFCs that are depicted in the right panel of Fig.~\ref{fig:q_dependence}. In stark contrast to the static density response, the STLS curves do not follow the PIMC data that we have obtained from Eq.~(\ref{eq:LFC}). While the former exhibit the same ordering as the latter for $q\lesssim3q_\textnormal{F}$, they exhibit a convergence towards a constant value in the limit of large $q$. In fact, this is a well known property of \emph{purely static theories of the LFC}~\cite{stls,Dornheim_PRL_2020_ESA,Dornheim_PRB_ESA_2021}, whereas the exact static limit of the dynamic LFC, $G(\mathbf{q}) = \lim_{\omega\to0}G(\mathbf{q},\omega)$ exhibits a more complicated behaviour. Indeed, Holas~\cite{holas_limit} has shown that the latter parabolically diverges for large $q$ at $T=0$, with the pre-factor being determined by the exchange--correlation contribution to the kinetic energy $K_\textnormal{xc}$~\cite{Militzer_Pollock_PRL_2002,Dornheim_PRB_nk_2021,Dornheim_PRE_2021,Hunger_PRE_2021}. Subsequently, based on highly accurate PIMC data, Dornheim \emph{et al.}~\cite{dornheim_ML,dornheim_electron_liquid} have discovered that the same qualitative trends persist also at finite temperatures. Remarkably, $K_\textnormal{xc}$ is known to attain \emph{negative values} for $T>0$ at some $r_{\mathrm{s}}$-$\theta$-combinations~\cite{Militzer_Pollock_PRL_2002, Hunger_PRE_2021}, which leads to a \emph{negative tail} of the LFC at large $q$; this is indeed the case for the total (spin symmetric) LFC, the spin antisymmetric LFC and the spin-diagonal LFC component at the present conditions. Moreover, we note that this negative slope seems to be exclusively caused by $G_{\mathrm{uu}}(\mathbf{q})$, whereas the spin-offdiagonal LFC appears to attain a constant value for large $q$. Unfortunately, this point cannot be conclusively settled on the basis of the present data due to the increasing Monte Carlo error bars. On the other hand, such a conclusion seems to be empirically supported by the almost non-existent imaginary-time dependence of $F_{\mathrm{du}}(\mathbf{q},\tau)$ shown in Fig.~\ref{fig:3D} above. Due to the substantially reduced magnitude of the imaginary-time diffusion between electrons of different spin-orientation, the corresponding intermediate scattering function behaves nearly classically. This will then be reflected in the behaviour of the LFC, which is known to converge towards a constant value at large $q$ in the classical limit~\cite{IIT}. Finally, in spite of the $\chi_{\mathrm{S}}(\mathbf{q})$, $\chi_{\mathrm{tot}}(\mathbf{q})$ convergence at $q\gtrsim2q_\textnormal{F}$, there is no corresponding $G_{-}(\mathbf{q})$, $G_{+}(\mathbf{q})$ convergence at short wavelengths, see the small difference in the constitutive Eqs.~(\ref{eq:density-density-paramagnetic}) and (\ref{eq:spin-spin-paramagnetic}) but also the non-zero large-$q$ limit of the spin-offdiagonal LFC.

Aiming to get a more direct practical insight into the spin-resolved density response of the warm dense UEG, we have also carried out PIMC simulations that are governed by the Hamiltonian of Eq.~(\ref{eq:Hamiltonian}) with $A_{\mathrm{u}}>0$ and with $A_{\mathrm{d}}=0$. In other words, we have only perturbed the spin-up electron species, but we have estimated how both spin-components react to the external perturbation. The results are shown in Fig.~\ref{fig:A_dependence}, where we show our PIMC data for the different induced densities for $q\approx0.84q_\textnormal{F}$ and $q\approx 2.52q_\textnormal{F}$. The green crosses have been obtained from earlier PIMC simulations~\cite{Dornheim_PRR_2021,Dornheim_PRL_2020} where both components were perturbed, and have been included as a reference. In addition, the red circles and blue diamonds show the actual spin-diagonal and spin-offdiagonal responses. The solid black lines show the corresponding predictions from linear-response theory (i.e., using our data for $\chi_{st}(\mathbf{q})$ that were obtained from the IT-ISFs), which are in excellent agreement to the raw PIMC data in all three cases in the vanishing perturbation limit. In addition, it is well known that the first nonlinear contribution to the density response at the wavenumber of the original perturbation (i.e., the first harmonic~\cite{Dornheim_PRR_2021}) is cubic in the perturbation amplitude $A$, and the dashed yellow lines have been obtained by fitting to the PIMC data the function~\cite{Dornheim_PRL_2020}
\begin{eqnarray}\label{eq:fit}
\rho_{st}(\mathbf{q},A) = \chi_{st}(\mathbf{q})A + \chi_{st}^{(\textnormal{cubic})}(\mathbf{q})A^3\ ,
\end{eqnarray}
where the linear and cubic response functions are the free parameters. It is straightforward that the inclusion of the cubic terms means that the analytical curves in Fig.~\ref{fig:A_dependence} remain valid for substantially stronger perturbation amplitudes. In addition, it should be pointed out that $ \chi_{st}^{(\textnormal{cubic})}(\mathbf{q})$ leads to a reduction of the actual density response compared to linear-response theory for both the diagonal and the offdiagonal density responses. In other words, nonlinear effects are actually more pronounced in the individual components, and they cancel to some degree in the full density response, cf.~the green crosses and the corresponding black and yellow curves. A detailed investigation of the spin-resolved nonlinear density response should be based on a corresponding generalization of the higher-order imaginary-time correlation functions introduced in Ref.~\cite{Dornheim_JCP_ITCF_2021}, which is beyond the scope of the present work.

In order to check the consistency of our PIMC implementation, we have included the linear coefficients from Eq.~(\ref{eq:fit}) as the yellow triangles in the left panel of Fig.~\ref{fig:q_dependence} for all the components. We find perfect agreement to the IT-ISF-based results for all three response functions, and for all three considered values of the wavenumber $q$.

A further perspective onto the physical meaning of the different $\chi_{st}(\mathbf{q})$ is provided in Fig.~\ref{fig:Density}, where we plot the density profile along the direction of the external perturbation in coordinate space. We consider the same wave numbers as in Fig.~\ref{fig:A_dependence}, and choose the perturbation amplitude $A=0.1$, where linear-response theory is accurate, although not exact; see the small deviations between the black curves and the red circles and green crosses in the left panel. In both cases, it can be clearly discerned that the perturbed spin-up electrons (red circles) strongly react to the externally imposed field and that they, on average, avoid the maximum of the latter. On the other hand, the directly unperturbed spin-down electrons (blue diamonds), exhibit the opposite trend at $q\approx0.84q_\textnormal{F}$ (left), but they hardly react for $q\approx2.52q_\textnormal{F}$ (right). As discussed earlier, at such wavenumbers, the full density response is nearly exclusively provided by the spin-diagonal component. 

\begin{figure}\centering
\includegraphics[width=0.475\textwidth]{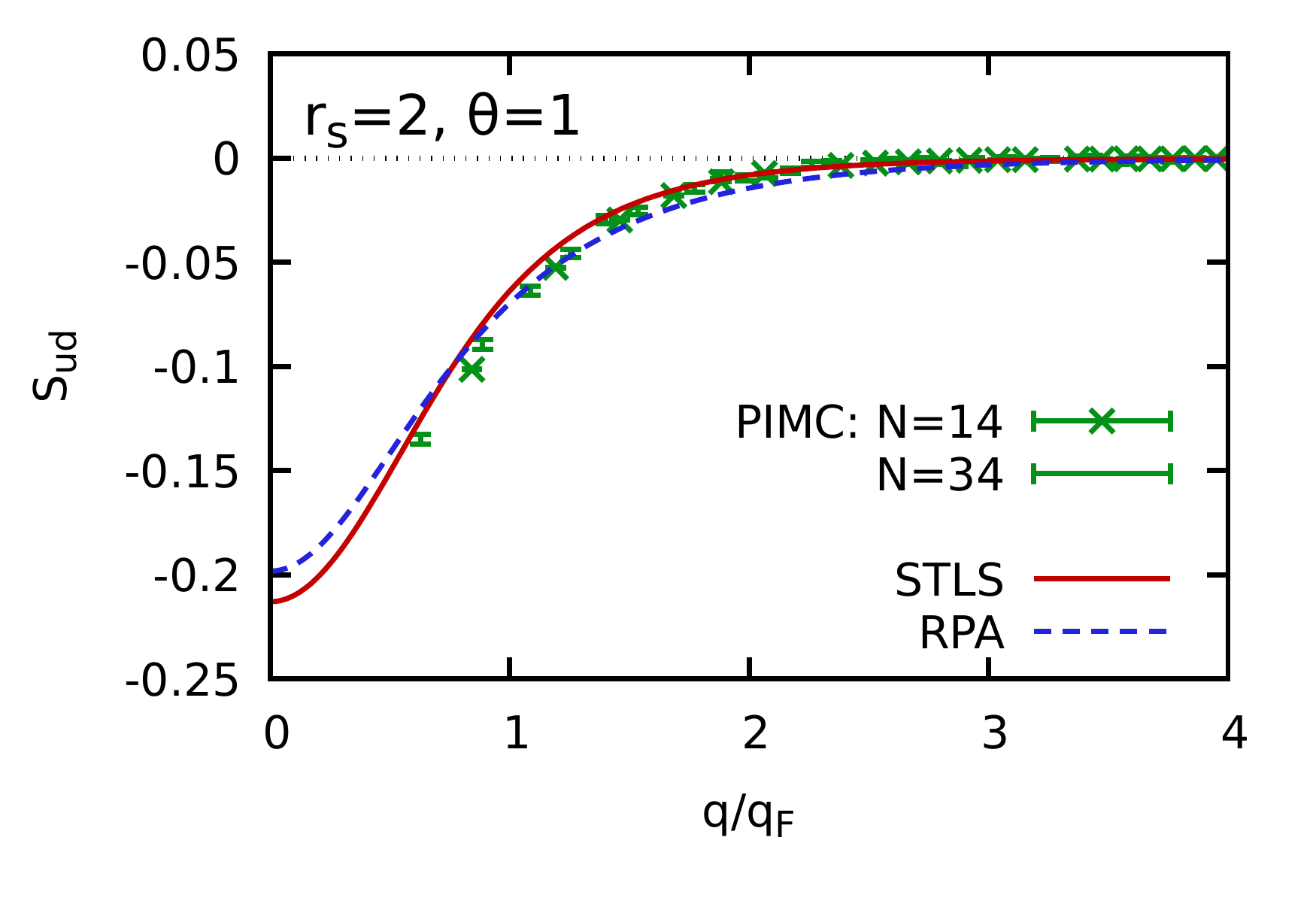}\\\vspace*{-1cm}
\includegraphics[width=0.475\textwidth]{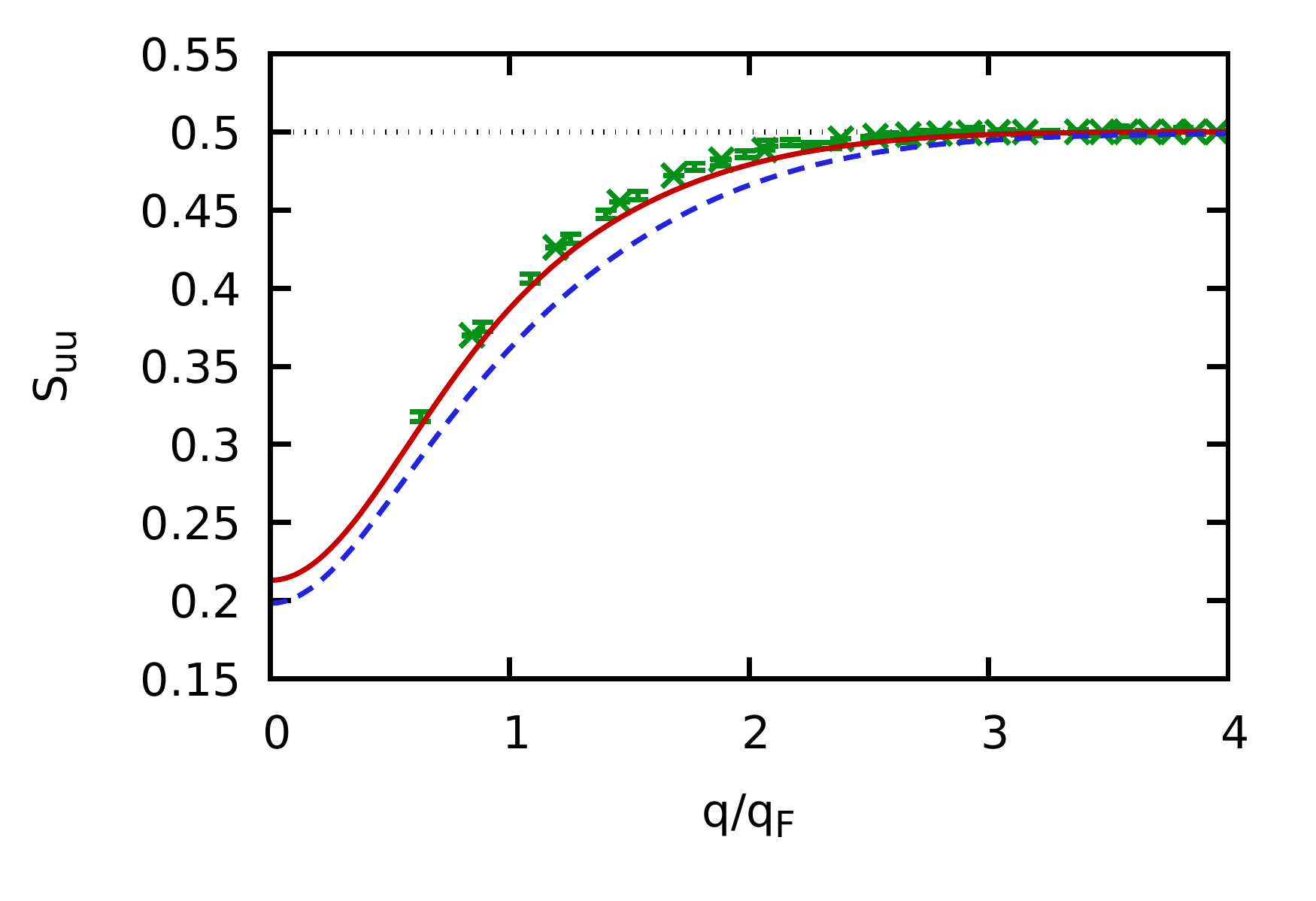}\\\vspace*{-1cm}
\includegraphics[width=0.475\textwidth]{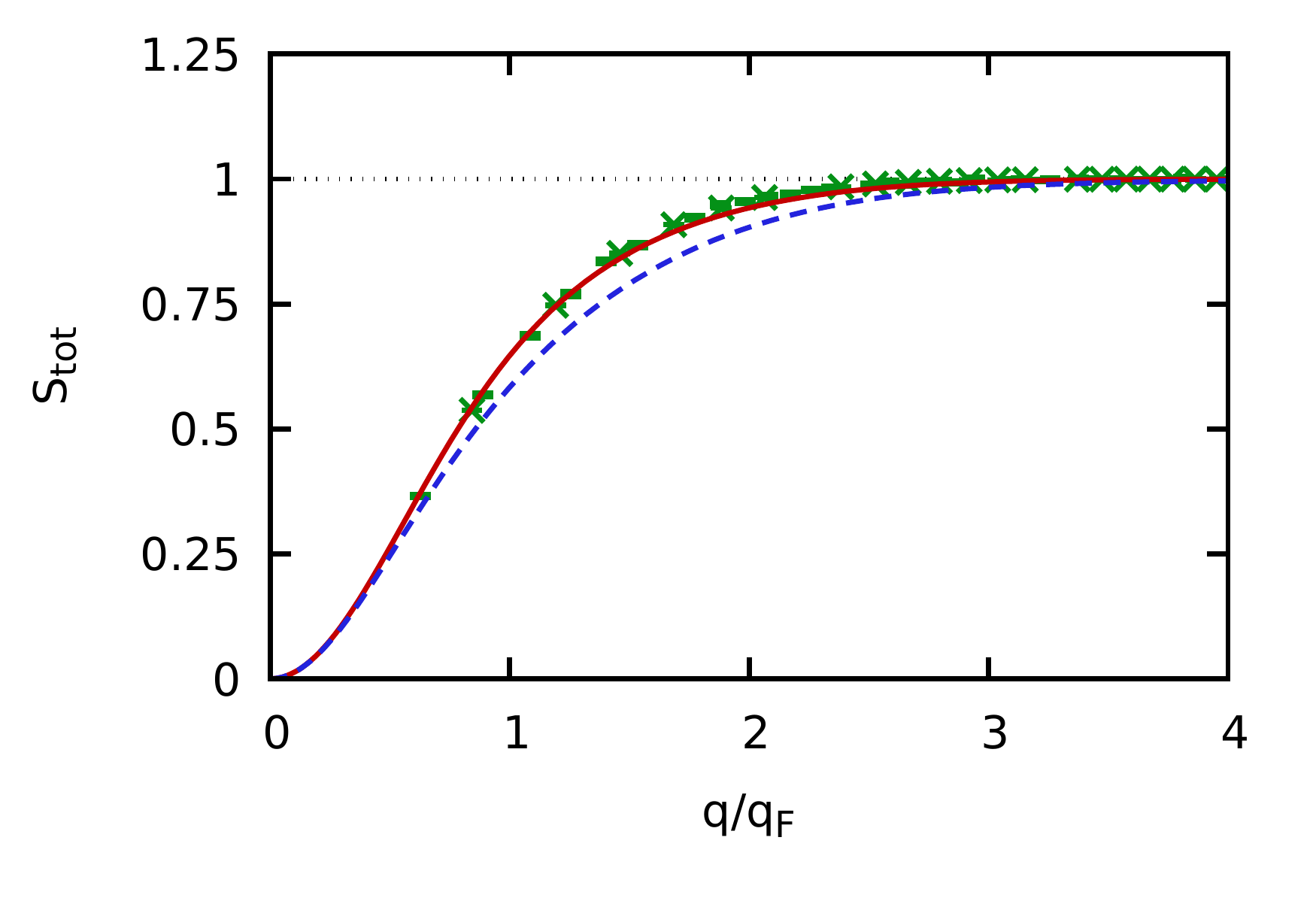}
\caption{\label{fig:SSF}
The spin-resolved components of the static structure factor [Eq.~(\ref{eq:S_st})] of the unpolarized UEG at $r_s=2$ and $\theta=1$. Green bars (crosses): PIMC results for $N=34$ ($N=14$); solid red line: finite-$T$ STLS scheme~\cite{stls,stls2}; dashed blue line: RPA. The top, center, and bottom panels show the spin-offdiagonal component $S_{\mathrm{ud}}(\mathbf{q})=S_{\mathrm{du}}(\mathbf{q})$, the spin-diagonal component $S_{\mathrm{uu}}(\mathbf{q})=S_{\mathrm{dd}}(\mathbf{q})$, and the total static structure factor $S_\textnormal{tot}(\mathbf{q})$, respectively.
}
\end{figure} 

\begin{figure*}\centering
\includegraphics[width=0.475\textwidth]{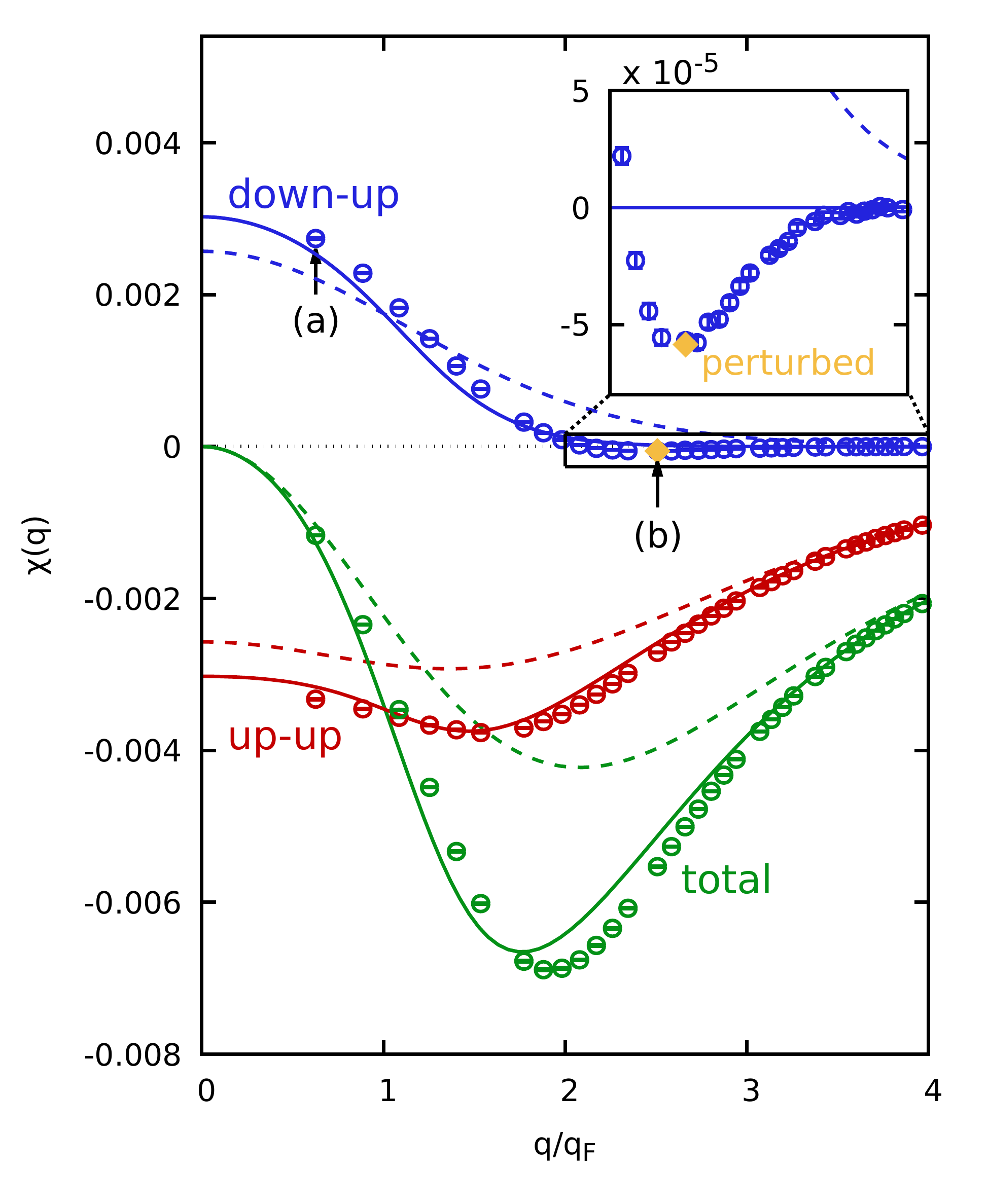}\includegraphics[width=0.485\textwidth]{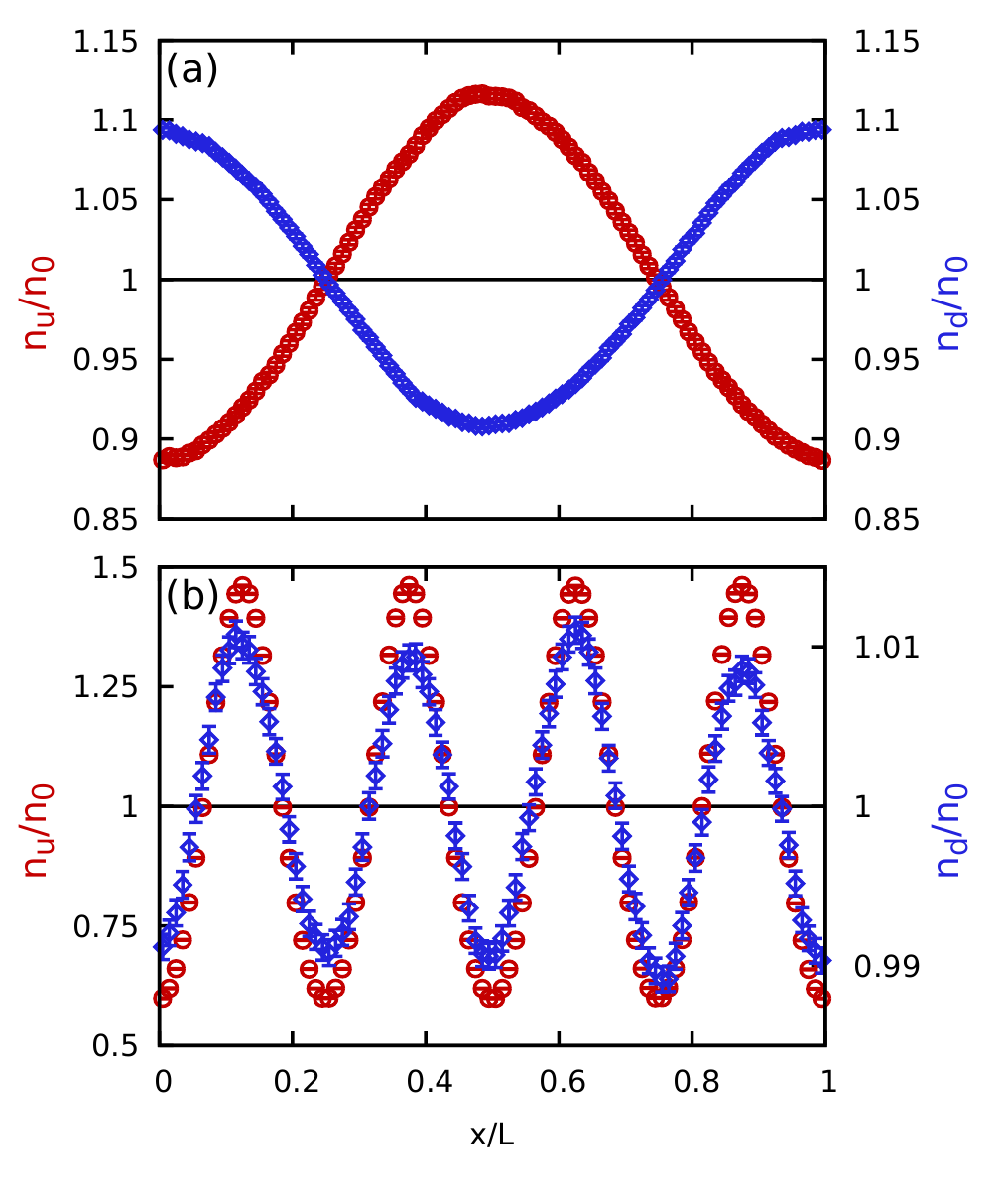}
\caption{\label{fig:attraction}
Left: spin-resolved density response functions $\chi_{\mathrm{uu}}(\mathbf{q})=\chi_{\mathrm{dd}}(\mathbf{q})$ (red) and $\chi_{\mathrm{du}}(\mathbf{q})=\chi_{\mathrm{ud}}(\mathbf{q})$ (blue), and the total density response function $\chi_\textnormal{tot}(\mathbf{q})$ (green) of the unpolarized UEG at $r_s=10$ and $\theta=1$. The points and solid (dashed) lines depict our new PIMC data and finite-$T$ STLS~\cite{stls,stls2} (RPA) results, respectively. Right: density profiles along the direction of the external perturbation with $q\approx 0.62q_\textnormal{F}$ and $A=0.002$ (a), $q\approx2.50q_\textnormal{F}$ and $A=0.01$ (b); the left and right $y$-axes correspond to the density of spin-up (red) and spin-down (blue) electrons. 
}
\end{figure*} 

Let us conclude our investigation of the spin-resolved density response of the warm dense UEG at the metallic density $r_{\mathrm{s}}=2$ by the analysis of the spin-resolved components of the static structure factor $S_{st}(\mathbf{q})$, cf.~Eq.~(\ref{eq:S_st}) in Sec.~\ref{sec:ITCF} above.
The respective results are shown in Fig.~\ref{fig:SSF}, where the spin-offdiagonal part $S_{\mathrm{du}}(\mathbf{q})=S_{\mathrm{ud}}(\mathbf{q})$, the spin-diagonal part $S_{\mathrm{uu}}(\mathbf{q})=S_{\mathrm{dd}}(\mathbf{q})$, and the total structure factor $S_\textnormal{tot}(\mathbf{q})$ are depicted in the top, center, and bottom panel. Specifically, the crosses and bars have been obtained for $N=14$ and $N=34$ unpolarized electrons, and no finite-size errors can be resolved within the given Monte Carlo error bars. In addition, the dashed blue and solid red lines show the corresponding results within the RPA and the STLS scheme. The latter exhibit the same qualitative trends as the PIMC reference data. We again observe that both these analytical approaches give the exact result for the total structure factor in the long-wavelength limit of $q\to0$~\cite{kugler_bounds},
\begin{eqnarray}
\lim_{q\to0}S_\textnormal{tot}(\mathbf{q}) = \frac{q^2}{2\omega_\textnormal{p}}\textnormal{coth}\left(\frac{\beta\omega_\textnormal{p}}{2}\right)\ ,
\end{eqnarray}
with $\omega_\textnormal{p}=\sqrt{3/r_s^3}$ the usual plasma frequency, but fail to give the exact long-wavelength results for the individual components $S_\textnormal{uu}(\mathbf{q})$ and $S_\textnormal{du}(\mathbf{q})$. Overall, as it is expected, the STLS scheme provides a substantially better agreement to the PIMC data over the entire $q$-range in all three cases. Nevertheless, it is evident that the impressive quality of the STLS results for the total structure factor $S_\textnormal{tot}(\mathbf{q})$ is the result of a fortunate error cancellation between the spin-diagonal and spin-offdiagonal components, where the STLS results are systematically too low and too high, respectively. The STLS discussion serves as a manifestation of how our new exact spin-resolved data for the density response can be exploited to give new insights into the performance of established dielectric schemes~\cite{stls,stls2,stolzmann,tanaka_hnc,arora,schweng}, and to guide the further development of novel schemes~\cite{Tolias_JCP_2021,castello2021classical}.

\subsection{Spin-resolved density response at the strongly coupled electron regime\label{sec:liquid}}

A further interesting research question is given by the dependence of the spin-resolved density response on the density parameter $r_s$. To this end, we have carried out additional PIMC simulations of the UEG at $\theta=1$ and $r_s=10$. These conditions are located at the margins of the strongly coupled electron liquid regime~\cite{dornheim_electron_liquid,quantum_theory} and exhibit a potential wealth of interesting physical effects. For example, Dornheim \emph{et al.}~\cite{dornheim_dynamic} have found a roton-like feature in the dispersion relation $\omega(q)$ of the dynamic structure factor $S(\mathbf{q},\omega)$ at these parameters, that has recently been explained by the spontaneous pair alignment of electrons~\cite{Dornheim_Nature_2022}. Furthermore, it has been reported that the effective interaction between two electrons in the presence of the UEG~\cite{Kukkonen_PRB_1979,Kukkonen_PRB_2021} becomes \emph{weakly attractive} at this density--temperature combination~\cite{Dornheim_Force_2022}.

In the left panel of Fig.~\ref{fig:attraction}, we show the corresponding spin-resolved components of the density response function, as obtained from our PIMC estimation of the respective $F_{st}(\mathbf{q},\tau)$. At a first glance, we perceive a qualitatively similar picture to the metallic density case shown in Fig.~\ref{fig:q_dependence}, \emph{i.e.}, the spin-diagonal and the spin-offdiagonal density response functions have opposite signs for $q\lesssim2q_\textnormal{F}$, and $\chi_{\mathrm{du}}(\mathbf{q})$ converges towards zero already around intermediate wave numbers $q/q_\textnormal{F}$. Furthermore, we observe that the RPA (dashed curves) is substantially less accurate in the present case. This is expected, given the increased coupling strength which leads to a more pronounced impact of the electronic exchange--correlation effects that are encoded into the LFCs $G_{st}(\mathbf{q})$ and are completely neglected within the RPA. Again, the STLS scheme constitutes a substantial improvement over the RPA, but does not correctly predict neither the magnitude nor the position of the peak in the total response function $\chi_\textnormal{tot}(\mathbf{q})$. When inspecting the spin-resolved components, we observe that the RPA systematically underestimates the true magnitude of $\chi_{\mathrm{uu}}(\mathbf{q})$ over the entire $q$-range, whereas it crosses the PIMC results for $\chi_{\mathrm{du}}(\mathbf{q})$ around $q=q_\textnormal{F}$. The STLS scheme, likewise, systematically underestimates the spin-diagonal component and crosses the PIMC data for $\chi_{\mathrm{du}}(\mathbf{q})$, but around $q=2q_\textnormal{F}$.

In addition to the above technical insight into the performance of different dielectric schemes, we detect a sign change in our PIMC data for the spin-offdiagonal density response function $\chi_{\mathrm{du}}(\mathbf{q})$ in the range $2q_\textnormal{F}\lesssim{q}\lesssim3q_\textnormal{F}$. This can be discerned particularly well in the inset showing a magnified segment around this remarkable feature. Aiming to further elucidate the physical origin and manifestation of this effect, we have carried out PIMC simulations of the UEG where only the spin-up electrons are subject to the external harmonic perturbation. The results for the spin-resolved density profiles along the direction of the perturbation are shown in the right column of Fig.~\ref{fig:attraction}, where the left and right ordinates correspond to the spin-up (red) and spin-down (blue) electrons, respectively. In particular, the top panel (a) has been obtained for $q\approx0.62q_\textnormal{F}$ and the perturbation amplitude $A_{\mathrm{u}}=0.002$; we re-iterate that there no perturbation is applied to the spin-down electrons, $A_{\mathrm{d}}=0$. Evidently, we observe a similar picture as for $r_{\mathrm{s}}=2$, cf.~Fig.~\ref{fig:Density} above, with the spin-up electrons escaping the external potential perturbation, and the spin-down electrons occupying the vacant space. Owing to the small wavenumber, the two contributions almost cancel, which indicates that the total electronic density response is small, as seen in the green curve in the left panel of Fig.~\ref{fig:attraction}. In panel (b), we have plotted the same information, but for $q\approx2.5q_\textnormal{F}$ and $A_{\mathrm{u}}=0.01$, which is located in the vicinity of the negative minimum of $\chi_{\mathrm{du}}(\mathbf{q})$. Remarkably, we find that the unperturbed spin-down electrons actually follow the perturbed spin-up electrons and, therefore, occupy the minima of the external potential acting on the latter. 

Before exploring the physical origin of this peculiar effect, we have performed one additional consistency check. More specifically, we have carried out a number of calculations at this wave number $q$ with different $A_u$, and have performed a cubic fit [cf.~Eq.~(\ref{eq:fit})] to the density response of the spin-down electrons $\rho_{\mathrm{d}}(\mathbf{q},A)$. The corresponding linear coefficient is depicted as the yellow diamond in the left panel of Fig.~\ref{fig:attraction}. Hence, the extracted density response function from the perturbation formalism is in perfect agreement to the data from the equilibrium IT-ISF formalism. The simulations of the perturbed system have thus conclusively shown that a) our IT-ISF implementation gives the correct linear-response prediction for this negative minimum and b) the peculiar simultaneous movement of both spin-components towards the minima of the external potential of the spin-up electrons is not an artefact of linear-response theory, but actually manifests in the true density profile shown in panel (b) without any expansions in powers of the perturbation amplitude.

\begin{figure}\centering
\includegraphics[width=0.475\textwidth]{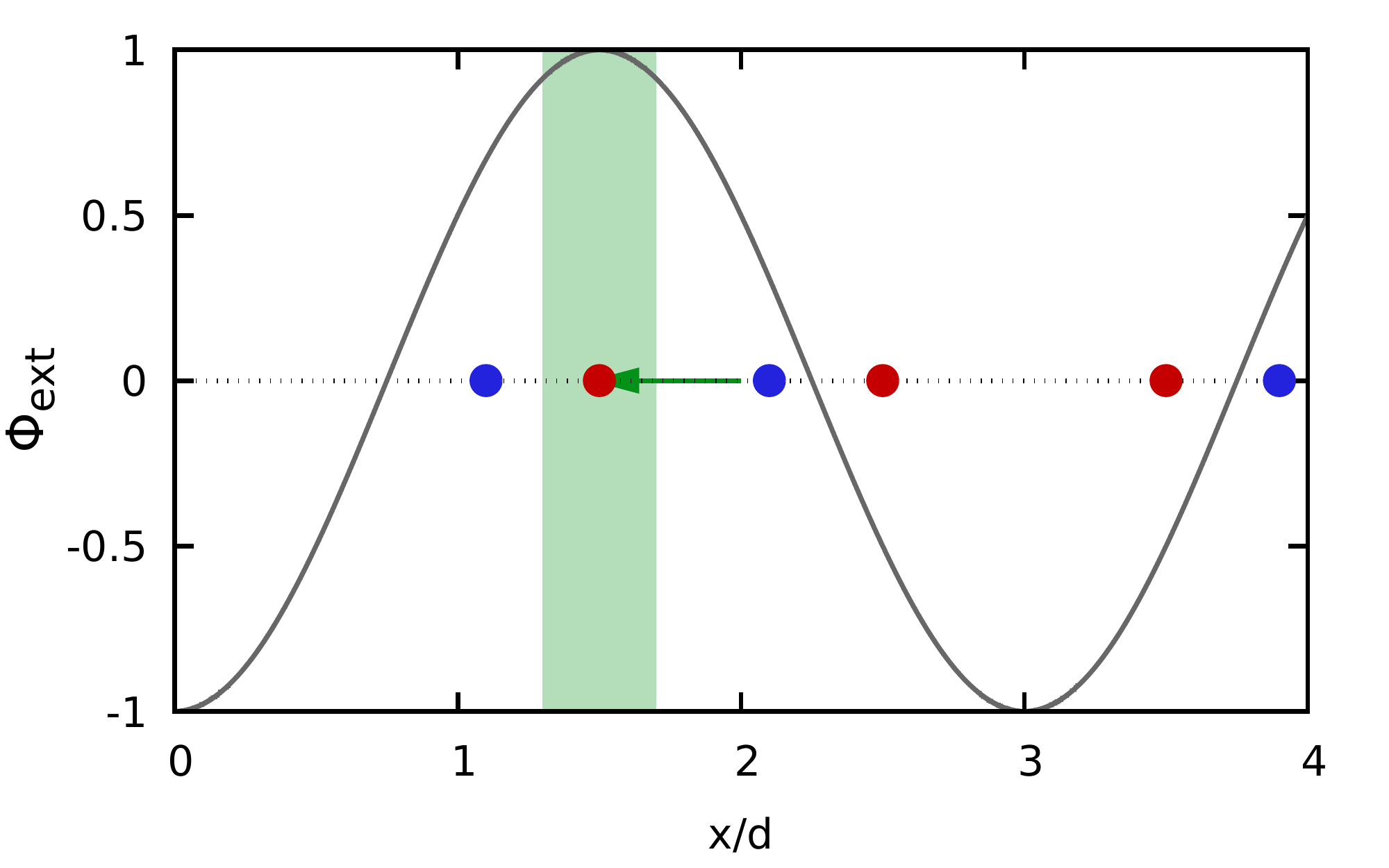}\\\vspace*{-0.9cm}\includegraphics[width=0.475\textwidth]{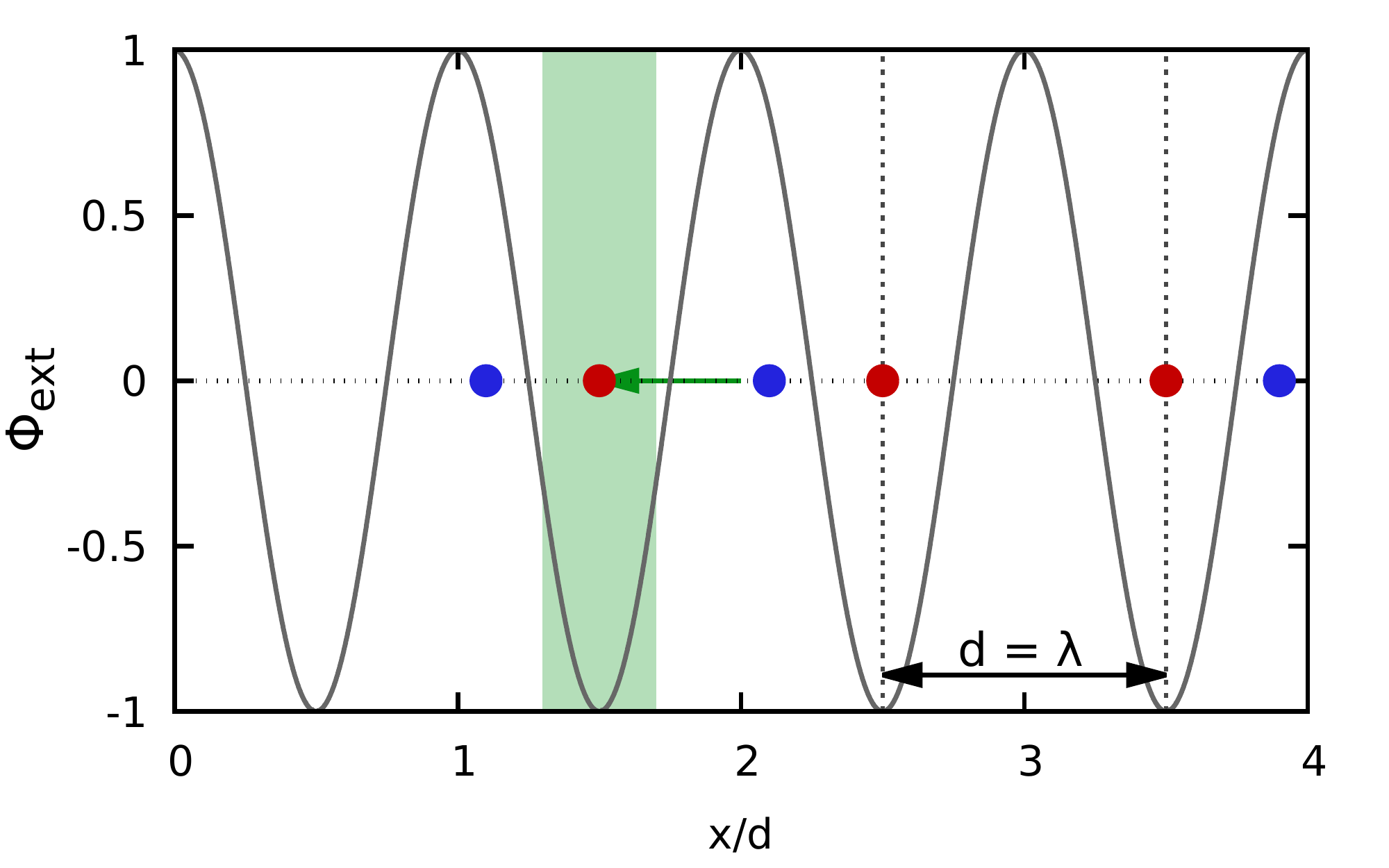}\\\vspace*{-0.9cm}\includegraphics[width=0.475\textwidth]{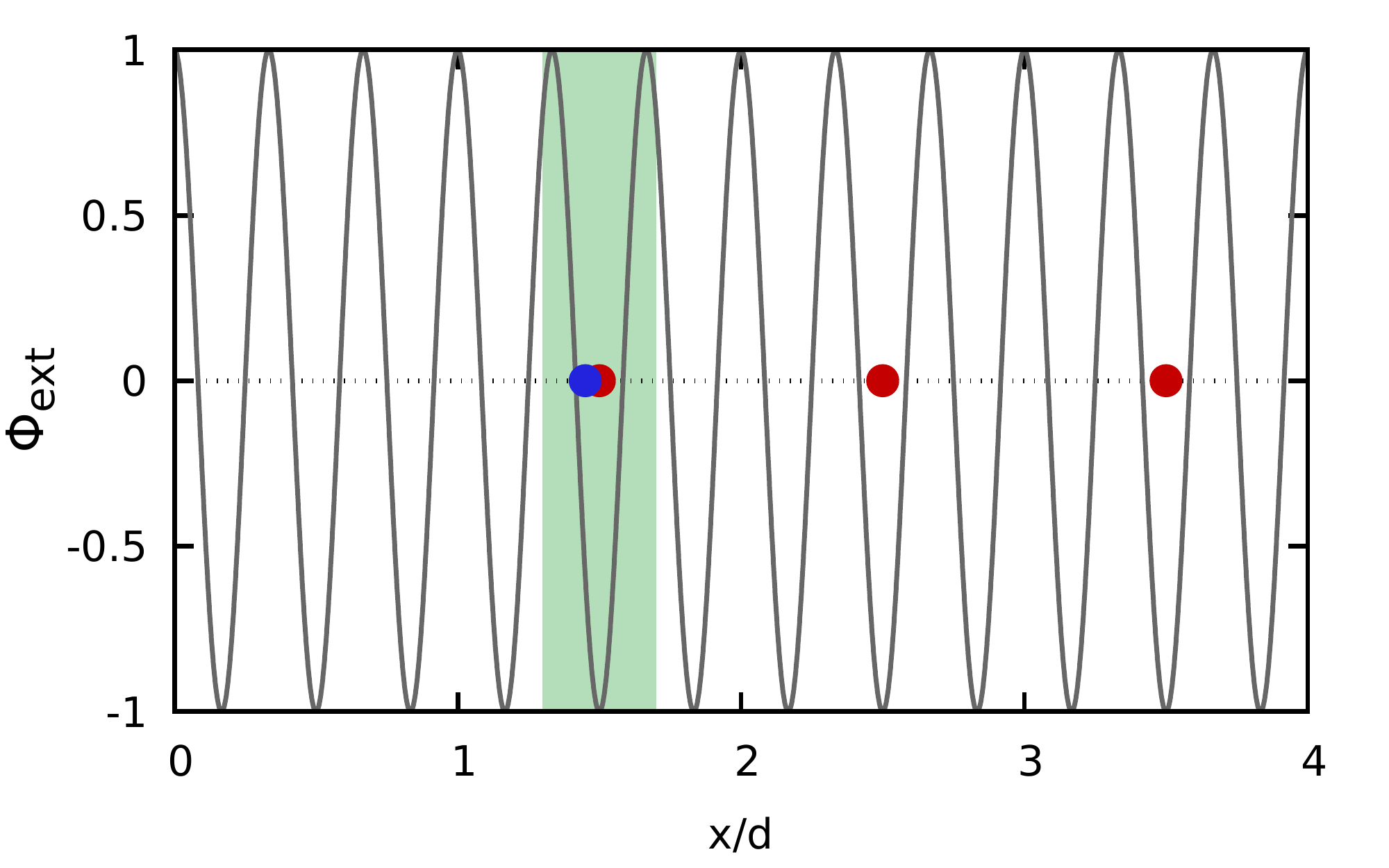}
\caption{\label{fig:spatial} An illustration of the length scales and the spatial structure of the harmonically perturbed UEG in the regime of \emph{effective attraction}, cf.~Fig.~\ref{fig:attraction}. The perturbed red electrons are, on average, aligned to the external potential applied directly to them (black sinusoidal lines). The unperturbed blue electron aligns itself to the minimum of the effective interaction energy landscape, which is depicted as the shaded green area. Top: long-wavelength regime ($\lambda>d$), both components react, but with the opposite sign; center: the perturbation is commensurate to the inter-particle distance ($\lambda\sim d$) and both components are aligned to the minima of $\Phi_\textnormal{ext}$; bottom: short-wavelength regime ($\lambda\ll r_{\mathrm{s}}$), only the spin-up electrons are aligned to $\Phi_\textnormal{ext}$.}
\end{figure} 

\begin{figure}\centering
\includegraphics[width=0.475\textwidth]{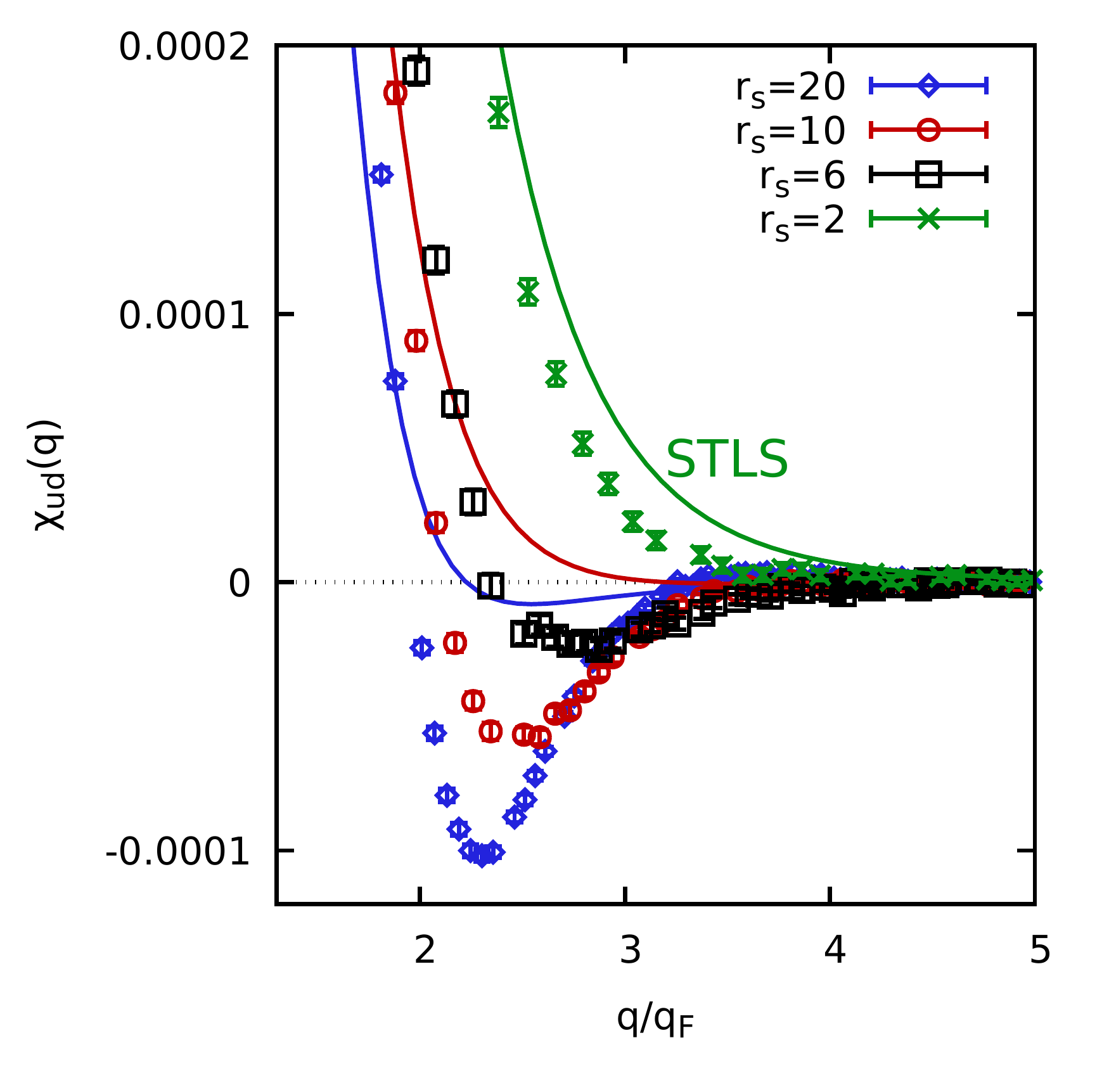}
\caption{\label{fig:zoom}
The spin-offdiagonal density response $\chi_{\mathrm{ud}}(\mathbf{q})$ of the unpolarized UEG at the electronic Fermi temperature $\theta=1$ for different values of the quantum coupling parameter $r_{\mathrm{s}}$. The solid curves show corresponding results of the finite-$T$ STLS scheme~\cite{stls,stls2}.
}
\end{figure} 

To understand the physical origin of this effect, we explore the spatial structure of the harmonically perturbed UEG, and the involved length scales, in Fig.~\ref{fig:spatial}. Here the red beads depict the perturbed spin-up electrons, which automatically move towards the minima of the external potential (black sinusoidal curves). In addition, the blue bead depicts a directly unperturbed spin-down electron. The top panel corresponds to the long wavelength regime, $\lambda\gg d$ (with $d\sim2r_s$ the average interparticle distance), and the blue spin-down electron will move towards the maximum of $\Phi_\textnormal{ext}(x)$, \emph{i.e.}, towards the minimum of its interaction energy landscape due to the other electrons that is indicated by the shaded grey area. 

The center panel corresponds to the regime of interest where $q\sim2.5q_\textnormal{F}$ and $\lambda\sim d$. In other words, the wavelength of the external potential is commensurate with the spatial structure of the system. Therefore, the blue spin-down electron is \emph{effectively pushed} by its neighbour to the right, and \emph{effectively attracted} by its neighbour to the left. The corresponding minimum of the interaction energy coincides with the minimum of $\Phi_\textnormal{ext}(x)$ in this case, which explains the \emph{negative sign} of the spin-offdiagonal density response function $\chi_{du}(\mathbf{q})$ in this regime.
Indeed, Dornheim \emph{et al.}~\cite{Dornheim_Force_2022} have recently reported such an effective attraction between two electrons in the UEG precisely in this regime, without any assumptions based on linear-response theory or other approximations. The observed behaviour of $\chi_{du}(\mathbf{q})$ thus constitutes a direct manifestation of electronic exchange--correlation effects. Consequently, it is not captured by either RPA or the STLS scheme, cf.~the inset in Fig.~\ref{fig:attraction}.

The bottom panel of Fig.~\ref{fig:spatial} schematically illustrates the spatial structure of the perturbed system in the single-particle regime where $\lambda\ll d$. While the spin-up electrons, as usual, move towards the minima of $\Phi_\textnormal{ext}(x)$, the unperturbed spin-down electrons can be anywhere in the vicinity of the shaded green area. Eventually, any correlations with the wavelength $\lambda$ will disappear, which means that $\chi_{\mathrm{du}}(\mathbf{q})=0$ in this regime.

In order to further explore the correlational origin of the negative minimum in $\chi_{\mathrm{du}}(\mathbf{q})$, we have carried out extensive PIMC simulations at the electronic Fermi temperature $\theta=1$ for four different values of the quantum coupling parameter $r_{\mathrm{s}}$. The results are depicted in Fig.~\ref{fig:zoom}, where the focus lies mainly on the wavenumber regime that is relevant for this effect. The data points correspond to our PIMC results for different $r_{\mathrm{s}}$, and the solid curves correspond to the STLS~\cite{stls,stls2} results for three representative densities. First and foremost, we observe that the magnitude of the effective attraction monotonically increases with $r_{\mathrm{s}}$, and starts to appear around $r_{\mathrm{s}}\gtrsim5$ at this value of $\theta$. It should be pointed out that this is in very close agreement to the emergence of the effective attraction between two electrons in the UEG that has been investigated in Ref.~\cite{Dornheim_Force_2022}. In addition, it is evident that the STLS scheme does not provide an adequate description of this effect; the shallow minimum in the solid blue curve for $r_s=20$ underestimates its true depth by an order of magnitude, and does not even qualitatively match the shape of the exact PIMC data.

\begin{figure}\centering
\includegraphics[width=0.475\textwidth]{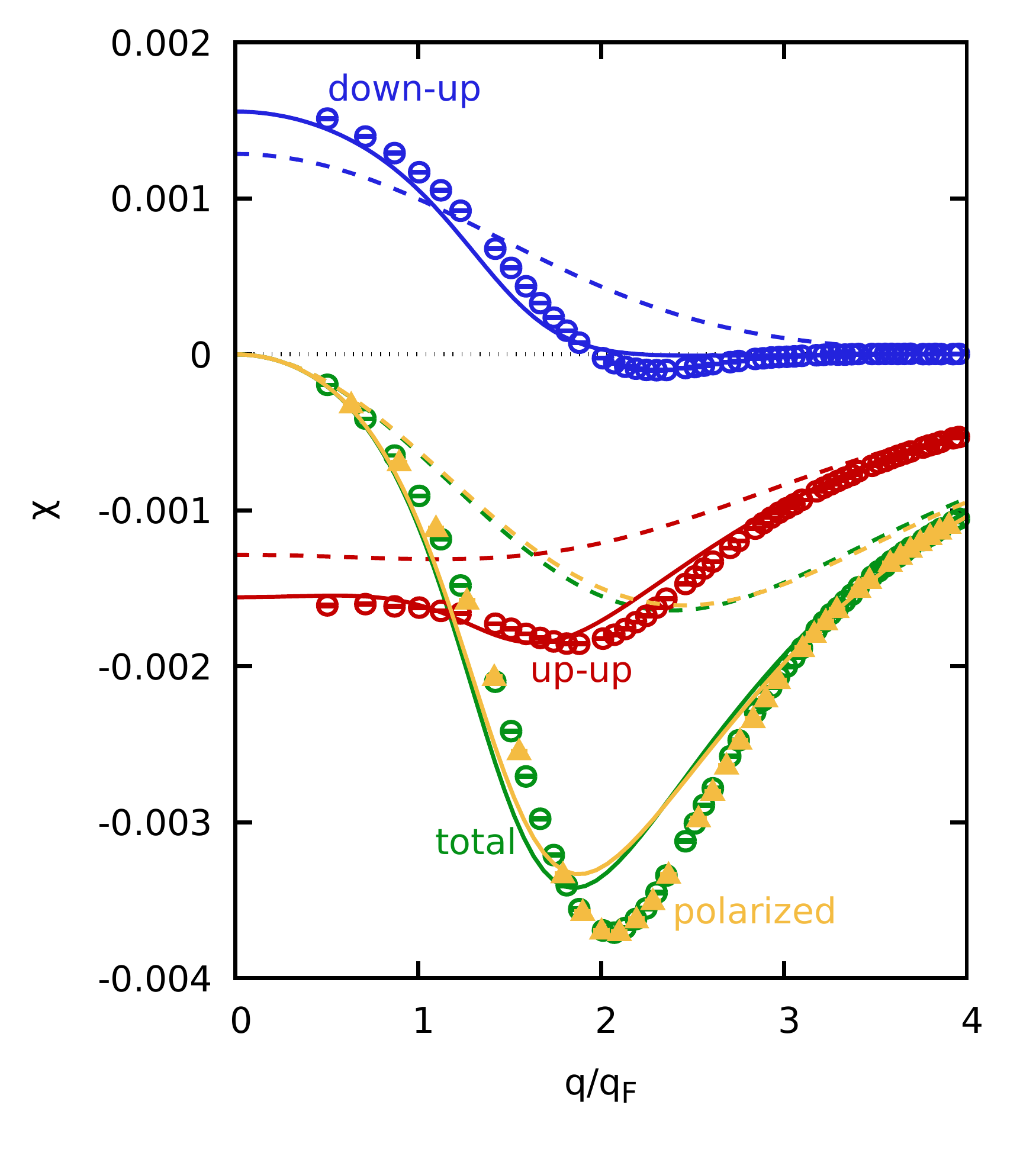}
\caption{\label{fig:rs20} The spin-resolved components of the static density response of the UEG at $\theta=1$ and $r_{\mathrm{s}}=20$. The red, blue, and green data sets show PIMC results for the spin-diagonal response $\chi_{\mathrm{uu}}(\mathbf{q})$, the spin-offdiagonal response $\chi_{\mathrm{ud}}(\mathbf{q})$, and the full response function $\chi_\textnormal{tot}(\mathbf{q})$, respectively. The yellow triangles data sets show PIMC results for the full response function of the spin-polarized ($\xi=1$) UEG at the same absolute $T$, $\beta=217.204$Ha$^{-1}$. The solid and dashed curves show the corresponding results of the finite-$T$ RPA and STLS scheme.}
\end{figure} 

\begin{figure*}\centering
\includegraphics[width=0.5\textwidth]{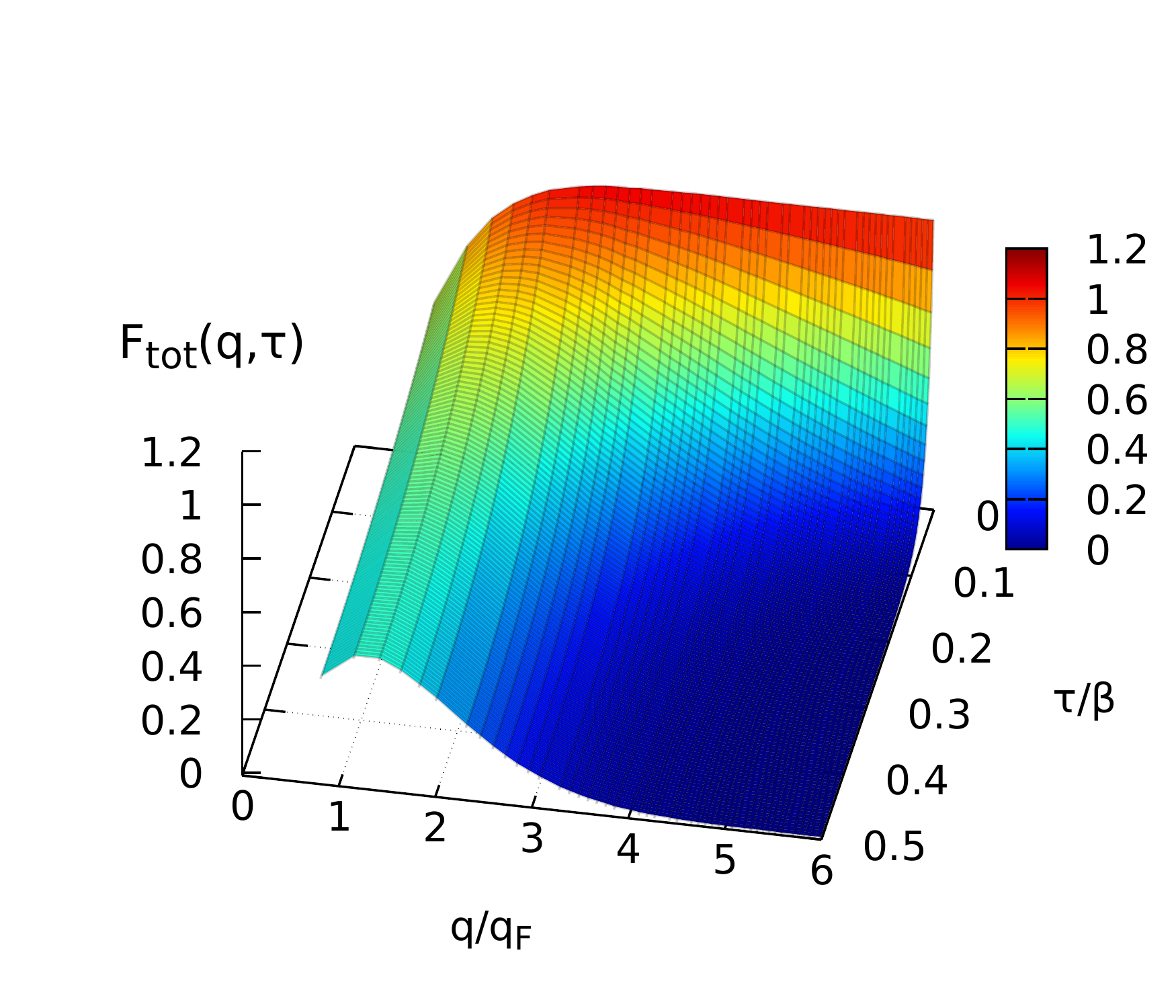}\includegraphics[width=0.5\textwidth]{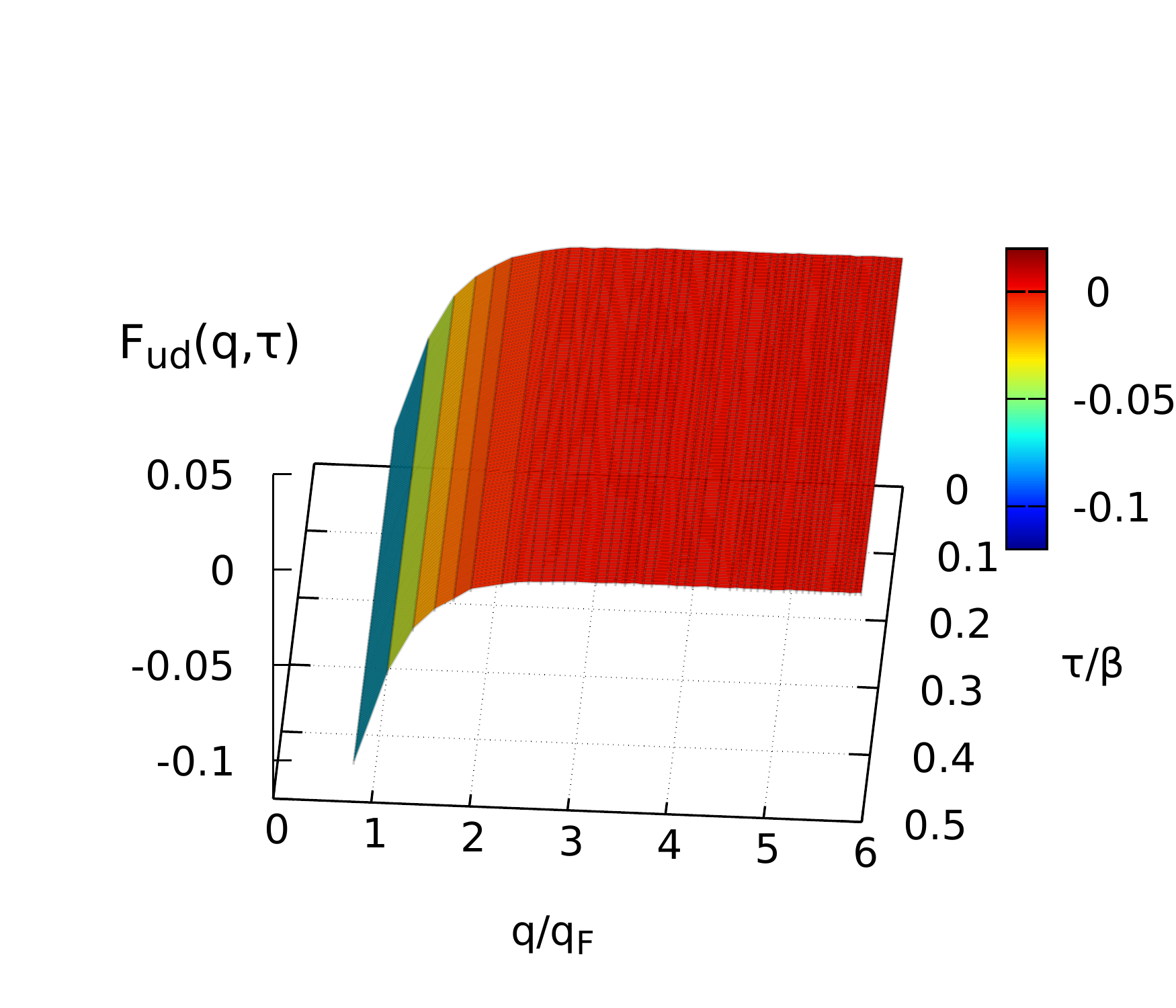}\\\vspace*{-1.18cm}\includegraphics[width=0.5\textwidth]{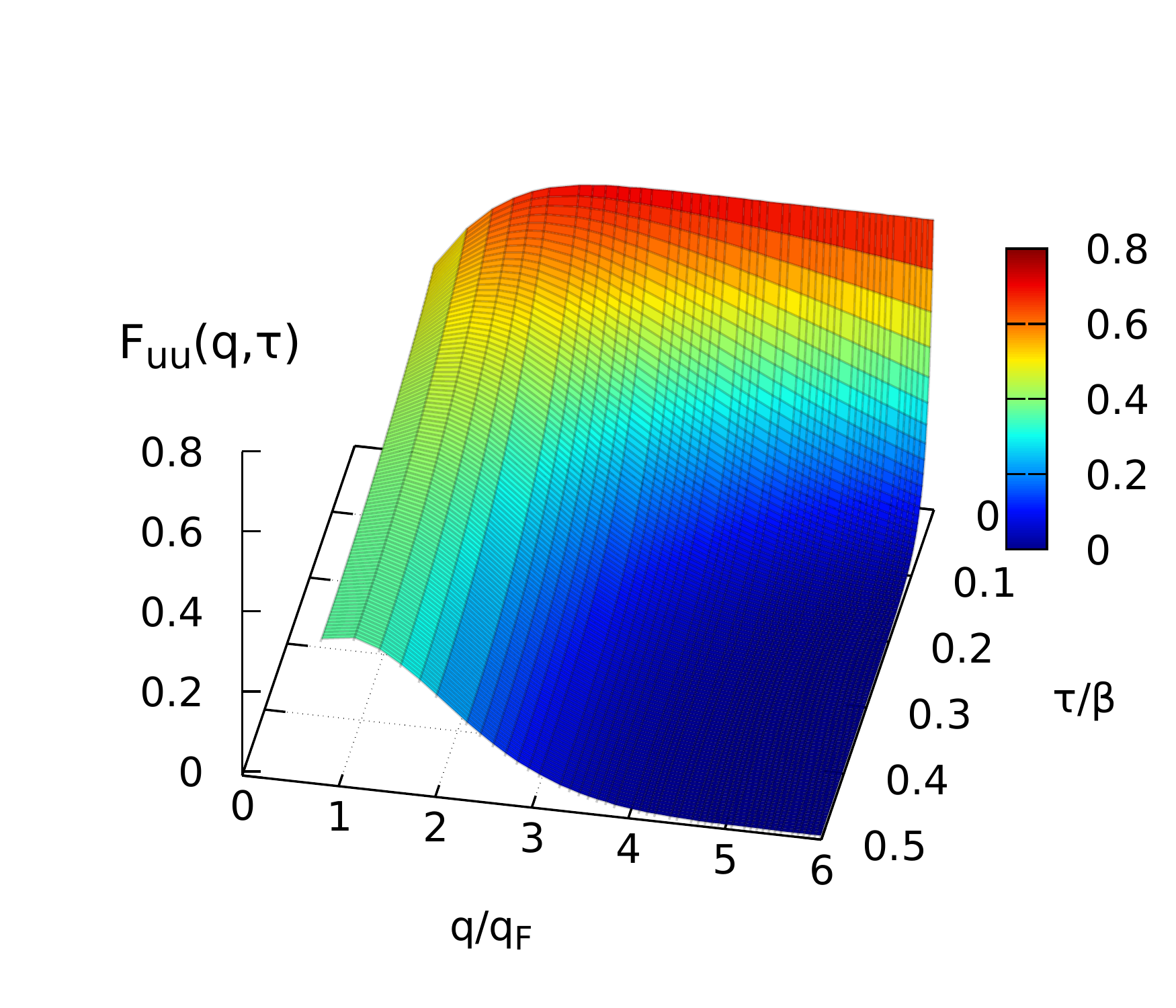}\includegraphics[width=0.5\textwidth]{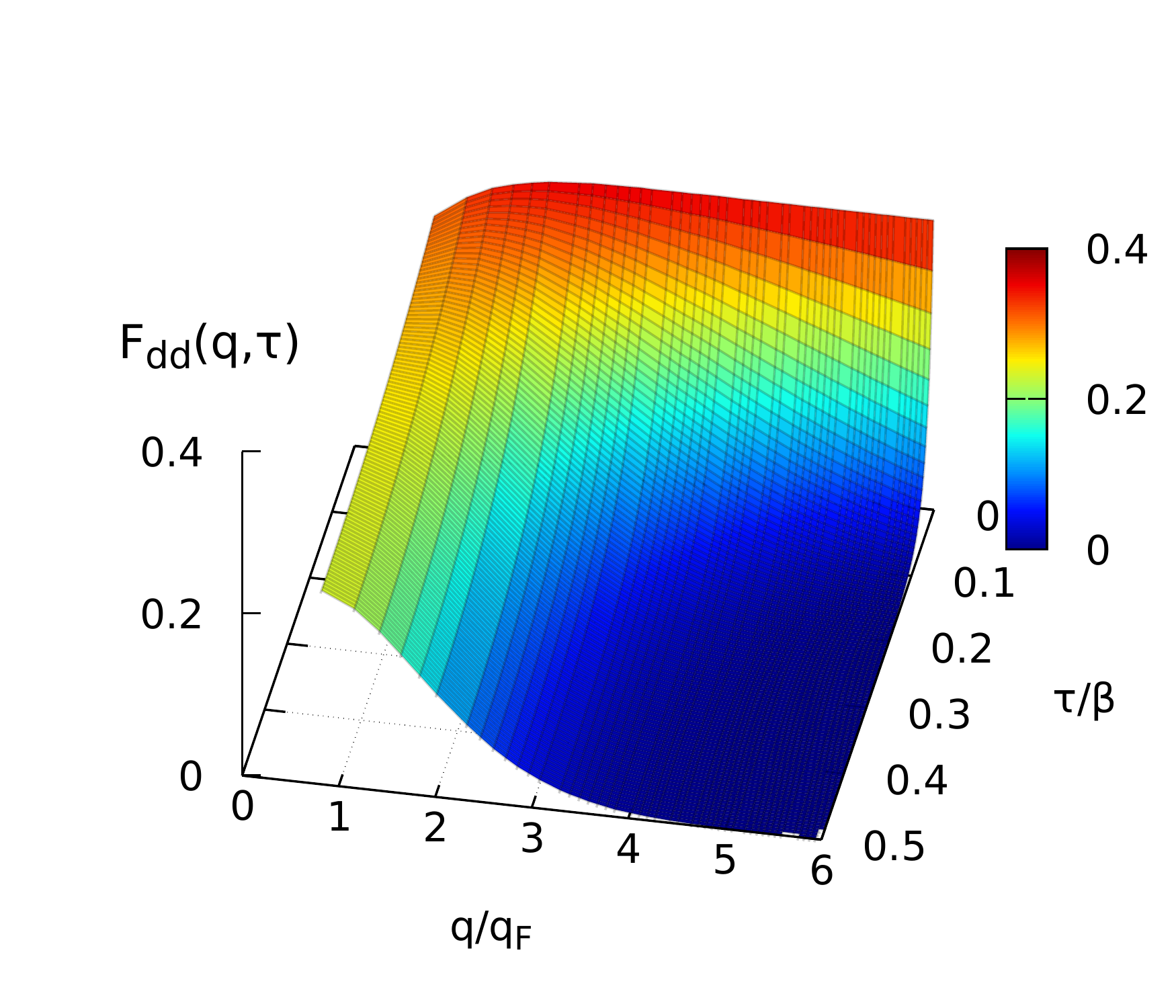}\vspace*{-0.49cm}
\caption{\label{fig:Xi_3D}
PIMC results for the spin-resolved components of the imaginary-time intermediate scattering function of the UEG at the intermediate polarization $\xi=1/3$, $r_s=2$ and $\theta=1$ (with the Fermi energy of the \emph{unpolarized} UEG as reference). Top left: the total IT-ISF $F_{\mathrm{tot}}(\mathbf{q},\tau)$. Top right: the spin-offdiagonal IT-ISF element $F_{\mathrm{ud}}(\mathbf{q},\tau)=F_{\mathrm{du}}(\mathbf{q},\tau)$. Bottom left: the spin-up diagonal IT-ISF element $F_{\mathrm{uu}}(\mathbf{q},\tau)$. Bottom right: the spin down diagonal IT-ISF element $F_{\mathrm{dd}}(\mathbf{q},\tau)$.
}
\end{figure*} 

We shall conclude the discussion of this effect with a note on the impact of the spin polarization parameter $\xi$. In fact, the spin-resolved density profiles that are illustrated in Figs.~\ref{fig:Density},\ref{fig:attraction} seem to imply that the observed attraction might be a spin-dependent phenomenon. To dispel this erroneous notion, we have performed PIMC simulations at $r_s=20$ and $\theta=1$, and the results are shown in Fig.~\ref{fig:rs20}. Specifically, the red, blue, green data sets correspond to the individual components $\chi_{\mathrm{uu}}(\mathbf{q})$, $\chi_{\mathrm{du}}(\mathbf{q})$, and $\chi_\textnormal{tot}(\mathbf{q})$ of the spin-unpolarized UEG ($\xi=0$). In addition, we have performed PIMC simulations of the \emph{spin-polarized} UEG (\emph{i.e.}, $\xi=1$ or $n=n_{\mathrm{u}}$, $n_{\mathrm{d}}=0$) at the same density and same absolute value of the inverse temperature $\beta=217.204$Ha$^{-1}$.
The results are shown as the yellow triangles in Fig.~\ref{fig:rs20}, and the PIMC data are in perfect agreement to the total density response function of the unpolarized system, i.e., the green circles. Clearly, the total density response function does not depend on the spin-polarization for such a large value of $r_s$. This is consistent to the recent investigation of the effective force and interaction potential between a pair of electrons in the UEG by Dornheim \emph{et al.}~\cite{Dornheim_Force_2022}, who have shown that any spin-dependence of these properties vanishes for distances $r\gtrsim r_s$. Therefore, the \emph{effective push} and \emph{effective attraction} that is experienced by the blue spin-down electron in the center panel of Fig.~\ref{fig:spatial} does not intrinsically depend on the spin-orientation of either electron. Spin-up electrons experience the same potential towards each other as with the spin-down electrons at the depicted distances. Instead, in this case, the apparent spin-dependence is entirely due to the spin-asymmetric external potential. Hence, the total density response does not depend on the spin-polarization and is the same for both the unpolarized and spin-polarized system as it can be clearly seen from the PIMC data points shown in Fig.~\ref{fig:rs20}.

Curiously, we do observe minor yet significant deviations between $\xi=0$ and $\xi=1$ in the analytical theories for $\chi_\textnormal{tot}(\mathbf{q})$ both within the RPA (dashed) and the STLS scheme (solid). In the real system, any dependence on the spin-polarization is suppressed by the strong electronic exchange--correlation effects that are exactly incorporated into the PIMC data. The analytical theories, on the other hand, are based on an approximate description which leads to the spurious, \emph{i.e.}, unphysical, observed spin-effect. 

\begin{figure*}\centering
\includegraphics[width=0.475\textwidth]{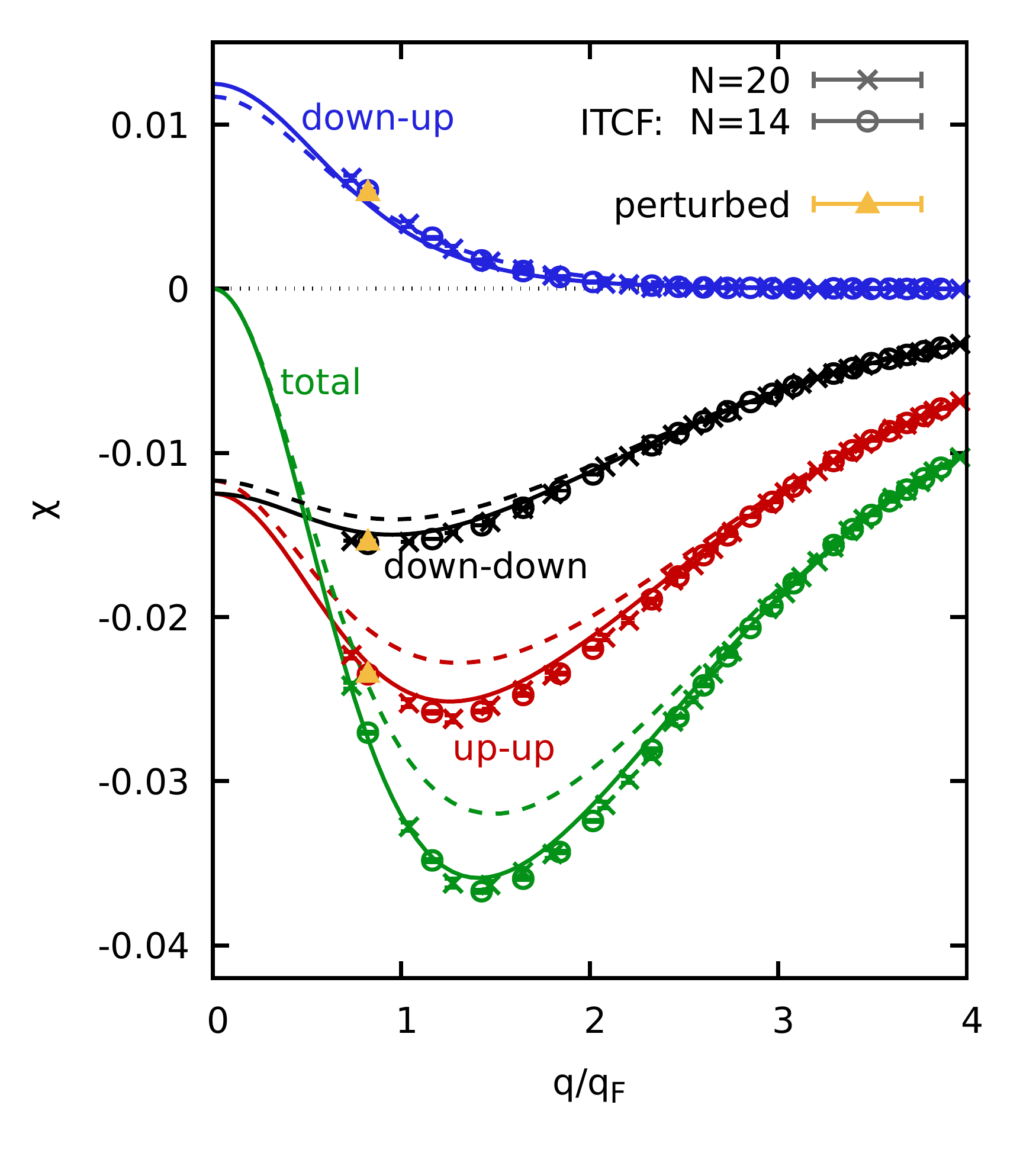}\includegraphics[width=0.475\textwidth]{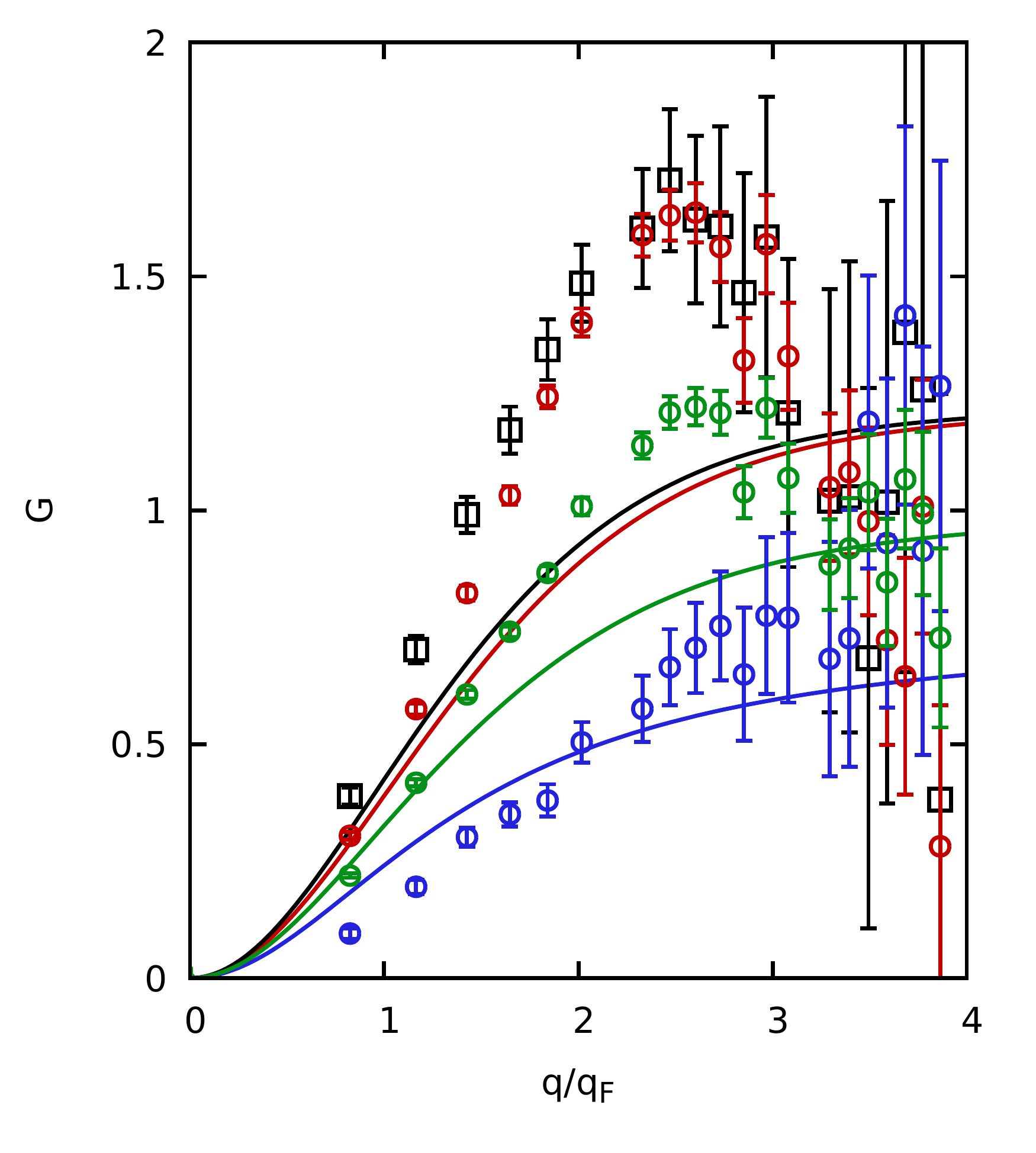}
\caption{\label{fig:Xi_chi}
Left: The spin-resolved components of the static density response of the partially polarized UEG at $\xi=1/3$, $r_{\mathrm{s}}=2$, $\theta=1$ (using the Fermi energy of the unpolarized UEG as a reference). Blue: the spin-offdiagonal component $\chi_{\mathrm{ud}}(\mathbf{q})=\chi_{\mathrm{du}}(\mathbf{q})$; red: the spin-diagonal component of the majority spin species $\chi_{\mathrm{uu}}(\mathbf{q})$; black: the spin-diagonal component of the minority spin species $\chi_{\mathrm{dd}}(\mathbf{q})$; green: the full density response $\chi_\textnormal{tot}(\mathbf{q}$). The yellow triangles correspond to the linear coefficient of a fit [cf.~Eq.~(\ref{eq:fit})] to the density response of the harmonically perturbed system, cf.~Fig.~\ref{fig:A_xi}. The crosses and the circles correspond to the PIMC results for $N=20$ and $N=15$ electrons. The solid and the dashed curves have been obtained within the finite-$T$ STLS scheme~\cite{stls,stls2} and the RPA. Right: The corresponding results for the spin resolved LFCs, see Eq.~(\ref{eq:LFC}).}
\end{figure*}

\subsection{Intermediate spin-polarizations\label{sec:intermediate}}

Let us conclude the investigation of the spin-resolved density response of the warm dense UEG by considering a case of intermediate polarization, $\xi=1/3$. In Fig.~\ref{fig:Xi_3D}, we show our PIMC results for the respective IT-ISFs at $r_{\mathrm{s}}=2$ and $\theta=1$ (using the Fermi energy of the unpolarized UEG as a reference). The top left panel shows the combined result of all electrons and closely resembles the corresponding $F_\textnormal{tot}(\mathbf{q},\tau)$ of the unpolarized UEG at the same conditions shown in Fig.~\ref{fig:3D} above. The same holds for the spin-offdiagonal element $F_{\mathrm{ud}}(\mathbf{q},\tau)=F_{\mathrm{du}}(\mathbf{q},\tau)$ that is illustrated in the top right panel. The bottom row shows our PIMC results for the spin-diagonal components, which are not equal for $\xi\neq0$. In fact, the IT-ISF of the majority component $F_{\mathrm{uu}}(\mathbf{q},\tau)$ shown in the bottom left panel exhibits a somewhat more pronounced structure than the minority component $F_{\mathrm{dd}}(\mathbf{q},\tau)$ shown on the right. Most likely, this can be traced back to a simple re-scaling of the respective $x$-axis. In fact, we divide the wave number $q$ by the Fermi wave number $q_\textnormal{F}$ of the unpolarized UEG at a total density of $r_{\mathrm{s}}=2$. For $\xi=1/3$, it holds $n_{\mathrm{u}} = 2n_{\mathrm{d}}$, such that $r_{\mathrm{s}}^{\mathrm{u}}<r_{\mathrm{s}}^{\mathrm{d}}$. Since the Fermi wave number scales as $q_\textnormal{F}\sim 1/r_{\mathrm{s}}$, the larger $r_{\mathrm{s}}^{\mathrm{d}}$ will mean that the results for $F_{\mathrm{dd}}(\mathbf{q},\tau)$ would effectively be shifted to the right, where $F_{\mathrm{uu}}(\mathbf{q},\tau)$, too, exhibits less structure, in particular for $\tau=\beta/2$.

We shall next examine the spin-resolved components of the static density response function, which are shown in the left panel of Fig.~\ref{fig:Xi_chi}. The comparison of the results for both the total density response $\chi_\textnormal{tot}(\mathbf{q})$ and the spin-offdiagonal response $\chi_{\mathrm{ud}}(\mathbf{q})=\chi_{\mathrm{du}}(\mathbf{q})$ to the corresponding results for $\xi=0$ shown in Fig.~\ref{fig:q_dependence} clearly illustrates that the overall effect of $\xi$ is small at these conditions; this observation holds both for the exact PIMC results and for the RPA (dashed) and STLS (solid) results. The situation becomes more interesting for the spin-diagonal density response, which differs between the majority and the minority spin. In the depicted $q$-range, the two curves only agree in the limit of $q\to0$, where they have to balance the respective limit of $\chi_{\mathrm{du}}(\mathbf{q})$, but they substantially disagree everywhere else. Overall, the density response of the spin-up electrons is systematically larger than the density response of the spin-down electrons due to their larger number density; yet, the simple ratio $n_{\mathrm{u}}/n_{\mathrm{d}}=2$ is \emph{not} reflected between $\chi_{\mathrm{dd}}(\mathbf{q})$ and $\chi_{\mathrm{uu}}(\mathbf{q})$. At this point, it should also be emphasized that, courtesy of the density imbalance, the normalized chemical potentials of the Fermi-Dirac distribution of each spin constituent are different. This would translate to differences that are not captured by the density ratio even in the non-interacting high density limit.  

In the right panel of Fig.~\ref{fig:Xi_chi}, we show the corresponding PIMC results for the spin resolved LFCs that have been extracted via Eq.~(\ref{eq:LFC}). Evidently, the Monte Carlo error bars are substantially larger compared to the LFC in the case of $\xi=0$ shown in Fig.~\ref{fig:q_dependence}. This is mainly a consequence of the fermion sign problem, which becomes more severe with increasing $\xi$ when the density and the (absolute) temperature are being kept constant. With increasing spin-polarization, fermionic exchange-cycles that can only form between electrons of the same spin-orientation can form in our PIMC simulations with increasing frequency; see Ref.~\cite{Dornheim_permutation_cycles} for a detailed discussion of the manifestation of permutation effects in the imaginary-time path integral picture. In addition, it should also be noted that the error bars of $G_{\mathrm{dd}}(\mathbf{q})$ and, to a lesser extend, $G_{\mathrm{ud}}(\mathbf{q})$ are comparably larger due to the reduced statistics on the spin-down electrons as $N_{\mathrm{d}}<N_{\mathrm{u}}$. 
Overall, the familiar ordering of the different spin components is recovered, with the spin-diagonal component of the minority electrons attaining the largest values. The same ordering appears in the STLS results (solid curves), although they do not resemble the true PIMC data even at a qualitative level.

\begin{figure}\centering
\includegraphics[width=0.475\textwidth]{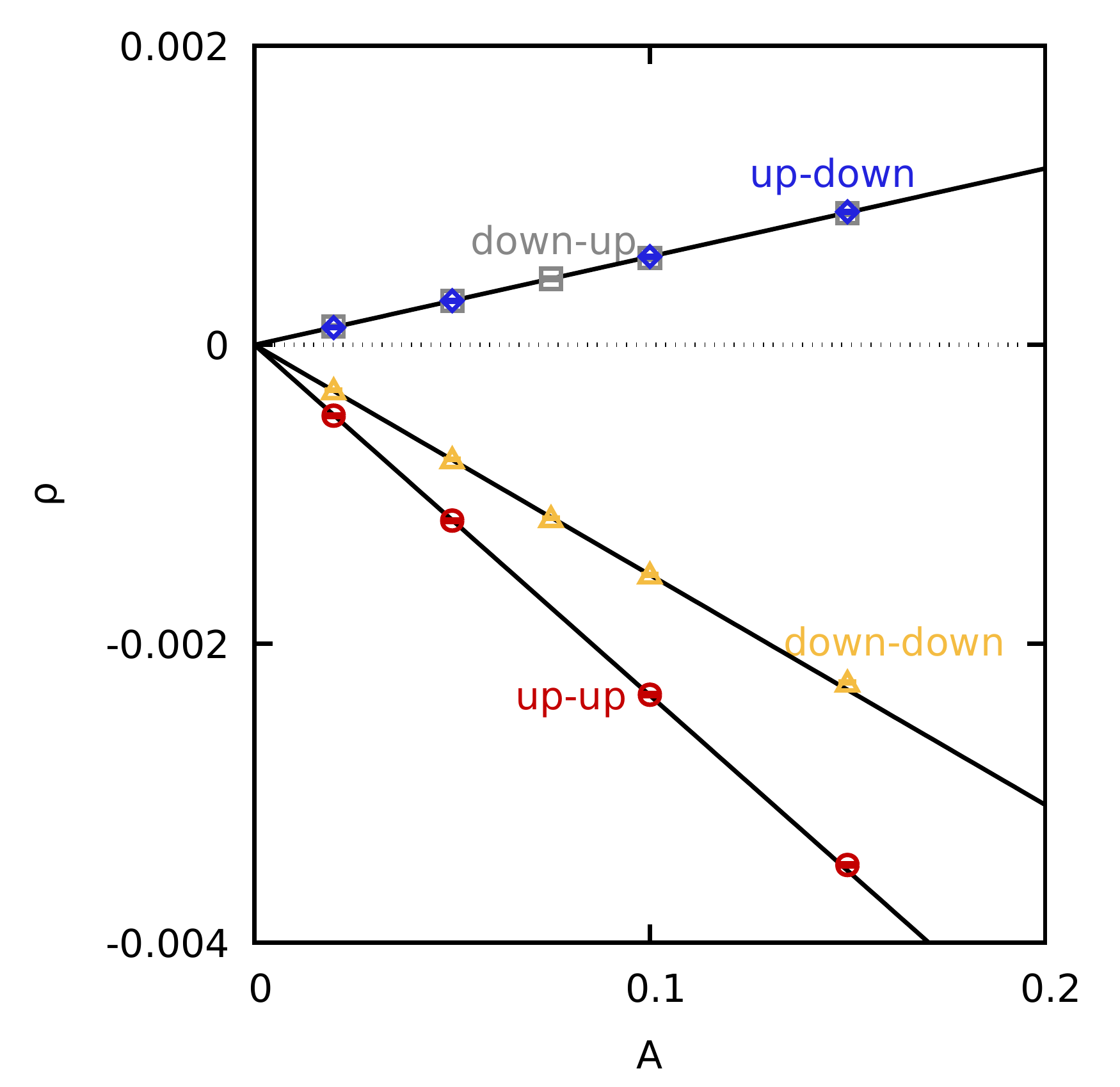}
\caption{\label{fig:A_xi}
Plot of the spin-resolved density response $\rho_s(\mathbf{q},A_t)$ versus the perturbation amplitude for the same conditions as in Fig.~\ref{fig:Xi_chi} ($r_s=2$, $\theta=1$, $\xi=1/3$) and at $q\approx0.82q_\textnormal{F}$.  Blue diamonds: the spin-offdiagonal response $\rho_{\mathrm{d}}(\mathbf{q},A_{\mathrm{u}})$; grey squares: the reciprocal spin-offdiagonal response $\rho_{\mathrm{u}}(\mathbf{q},A_{\mathrm{d}})$; red circles: the spin-diagonal response of the majority spin species $\rho_{\mathrm{u}}(\mathbf{q},A_{\mathrm{u}})$; yellow triangles: the spin-diagonal response of the minority spin species $\rho_{\mathrm{d}}(\mathbf{q},A_{\mathrm{d}})$. The solid black line corresponds to the prediction of linear-response theory.
}
\end{figure} 

As the final research question to be investigated in this work, we have performed PIMC simulations of the harmonically perturbed system with $\xi=1/3$. The results for the density response of the different components as function of the perturbation amplitude of the spin-up electrons $A_{\mathrm{u}}$ are shown in Fig.~\ref{fig:A_xi}; the spin-down electrons are not subject to an external potential, \emph{i.e.}, $A_{\mathrm{d}}\equiv0$. First and foremost, we note that linear-response theory (black lines) is accurate for all components over the entire investigated range of the amplitude. In addition, the spin-up electrons (red circles) exhibit a density response that is larger in magnitude compared to that of the spin-down electrons (yellow triangles), as it is expected from Fig.~\ref{fig:Xi_chi}. Regarding the spin-offdiagonal component, the blue diamonds show $\chi_{\mathrm{du}}(\mathbf{q})$; the density response of the spin-down electrons due to the external perturbation of the spin-up electrons. Furthermore, we have performed PIMC simulations where we have perturbed the spin-down electrons, and, instead, we have set $A_{\mathrm{u}}\equiv0$; the corresponding spin-offdiagonal density response $\rho_{\mathrm{u}}(\mathbf{q},A_{\mathrm{d}})$ is depicted by the grey squares, which are in perfect agreement to the reciprocal $\rho_{\mathrm{d}}(\mathbf{q},A_{\mathrm{u}})$. The validity of the general relation $\chi_{\mathrm{ud}}(\mathbf{q})\equiv\chi_{\mathrm{du}}(\mathbf{q})$, that should also hold in the case of intermediate polarizations, is yet another manifestation of the internal consistency of our PIMC simulations.

\section{Summary and Discussion\label{sec:summary}}

In this work, we have presented the first highly accurate \emph{ab initio} PIMC results for the spin-resolved density response of the warm dense UEG. This has been achieved via different routes: a) the estimation of spin-resolved imaginary-time intermediate scattering functions from unperturbed simulations that makes it possible to obtain the full wavenumber dependence of the spin-resolved density response from a single simulation of the unperturbed system; b) the direct estimation of the spin-resolved density response at selected wavenumbers from simulations where one spin-component is subject to an external harmonic potential whereas the other remains unperturbed, which also makes it possible to study non-linear effects. As expected, both procedures are in perfect agreement with each other at the appropriate limit of infinitesimal perturbation amplitudes. Furthermore, we have utilized our new exact spin-resolved density response data in order to extract the spin-resolved components of the static local field correction, which contains the full, wavenumber resolved information about electronic exchange--correlation effects. 

In addition to their fundamental value, our unique cutting-edge results allow us to assess the accuracy of previously widely-used approximations. The popular RPA gives the correct qualitative, though not quantitative, description at the metallic density of $r_{\mathrm{s}}=2$, but it completely breaks down from the threshold of strong coupling ($r_{\mathrm{s}}=10$), as it is expected. In contrast, the finite-$T$ STLS formalism~\cite{stls,stls2} gives accurate results both for the total density response function and the total static structure factor at $r_s=2$. Interestingly, this has been identified to be the consequence of a fortunate error cancellation as the spin-resolved components are significantly less accurate for both $\chi(\mathbf{q})$ and $S(\mathbf{q})$. Furthermore, the static LFC substantially deviates from our PIMC results. At $r_{\mathrm{s}}=10$, the accuracy of the STLS, too, deteriorates, and it also exhibits qualitative disagreements.

From a physical perspective, our PIMC-based study has given new microscopic insights into the interplay of the spin-resolved components of the static density response in the warm dense UEG. Typically, the perturbed component reacts by aligning to the minima of the external potential, which implies a \emph{negative response function}. The unperturbed component, on the other hand, then, on average, occupies this vacant space and will predominantly be located around the maxima of $\Phi_\textnormal{ext}(r)$. This means that their response function has the opposite sign of the perturbed component, \emph{i.e.}, it is positive. Remarkably, we find that the unperturbed electrons actually follow the perturbed component for $2q_\textnormal{F}\lesssim q \lesssim 3q_\textnormal{F}$ at strong coupling. This can be directly traced back to the fact that the associated wavelength $\lambda$ is comparable to the mean interparticle distance in this case. In other words, the unperturbed component is effectively moved by the Coulomb correlations to an unoccupied minimum of the external potential. Evidently, this remarkable effect is directly connected to the effective attraction between two electrons in the UEG that has been discussed in the recent Ref.~\cite{Dornheim_Force_2022}.

Finally, our exact PIMC simulations have given us straightforward access to correlations in the imaginary-time domain. Remarkably, we have found that the spin-offdiagonal components of the imaginary-time intermediate scattering function $F_\mathrm{ud}(\mathbf{q},\tau)$ hardly depend on $\tau$. Specifically, the impact of imaginary-time diffusion on particles of different species is orders of magnitude smaller compared to the diagonal component $F_\mathrm{ss}(\mathbf{q},\tau)$, with the latter mainly being driven by thermal self-diffusion.

We are convinced that our study opens up a number of possibilities for future investigations. i) Accurate results for the spin-resolved density response can be used to benchmark existing approximations and may guide the further development of new, improved dielectric theories~\cite{arora,tanaka_hnc,castello2021classical,Tolias_JCP_2021, doi:10.1063/5.0062325, jctc_22}; ii) Our new insights into the interplay of the spin-resolved components of the density response will constitute an important basis for future investigations of matter in an external magnetic field~\cite{Haensel}; iii) The practical manifestation of the effective electron--electron attraction~\cite{Dornheim_Force_2022} in the UEG density response deserves further exploration. This might include the study of \emph{spin-resolved pair alignment}~\cite{Dornheim_Nature_2022}, the possible formation of a charge- and spin-density-wave~\cite{quantum_theory,Overhauser_PhysRev_1962}, and its potential connection to superconductivity~\cite{Takada_PRB_1993}.

\section*{Acknowledgments}
This work was partially supported by the Center for Advanced Systems Understanding (CASUS) which is financed by Germany’s Federal Ministry of Education and Research (BMBF) and by the Saxon state government out of the State budget approved by the Saxon State Parliament.
The PIMC calculations were carried out at the Norddeutscher Verbund f\"ur Hoch- und H\"ochstleistungsrechnen (HLRN) under grant shp00026, and on a Bull Cluster at the Center for Information Services and High Performance Computing (ZIH) at Technische Universit\"at Dresden.

\bibliography{bibliography}

\begin{thebibliography}{139}
\bibitem [{rep()}]{repo}%
  \BibitemOpen
  \href@noop {} {}\bibinfo {note} {A link to a repository containing all PIMC
  raw data will be made available upon publication.}\BibitemShut {Stop}%
\end{thebibliography}
\end{document}